\documentclass[a4paper]{article}
\usepackage[a4paper,margin=2.4cm]{geometry}

\usepackage{multirow}
\usepackage{times}
\usepackage[T1]{fontenc}
\usepackage{caption}
\usepackage{graphicx}
\expandafter\let\csname equation*\endcsname\relax 
\expandafter\let\csname endequation*\endcsname\relax 
\usepackage{amsmath}
\usepackage{amssymb}
\numberwithin{equation}{section}
\numberwithin{figure}{section}
\usepackage{tikz}
\usetikzlibrary{shapes}
\usepackage{tikz-3dplot}

\usepackage[hidelinks]{hyperref}
\hypersetup{
 pdfauthor={Nathan Clisby},
 pdftitle={Endless self-avoiding walks},
 pdfsubject={},
 pdfkeywords={self-avoiding walk, critical exponent, enumeration, critical amplitude},
 colorlinks = false,
 pdfborder = {0 0 0}, 
 }

\newcommand{\mue}{\ensuremath{\mu_{\rm e}}}
\newcommand{\Ae}{\ensuremath{A_{\rm e}}}
\newcommand{\Be}{\ensuremath{B_{\rm e}}}
\newcommand{\Ce}{\ensuremath{C_{\rm e}}}
\newcommand{\De}{\ensuremath{D_{\rm e}}}
\newcommand{\gammae}{\ensuremath{\gamma_{\rm e}}}
\newcommand{\nue}{\ensuremath{\nu_{\rm e}}}

\newcommand{\gammaek}{\ensuremath{\gamma_{{\rm e},k}}}
\newcommand{\nuek}{\ensuremath{\nu_{{\rm e},k}}}

\newcommand{\gammaen}{\ensuremath{\gamma_{{\rm e},n}}}

\DeclareFontFamily{OT1}{pzc}{}
\DeclareFontShape{OT1}{pzc}{m}{it}{<-> s * [1.10] pzcmi7t}{}
\DeclareMathAlphabet{\mathpzc}{OT1}{pzc}{m}{it}
\newcommand{\bridge}{\ensuremath{\mathpzc{b}}}

\usepackage{placeins}

\newcommand{\Z}{\ensuremath{\mathbb{Z}}}
\newcommand{\wex}{\ensuremath{\omega_{\rm ex}}}

\def\root{\draw[fill] +(0,0) circle (5pt);}

\def\sww{\draw ++(0,0)}

\def\sdw{\draw[style=densely dotted] ++(0,0)}
\def\ew{;}
\def\r{-- ++(1,0)}
\def\l{-- ++(-1,0)}
\def\u{-- ++(0,1)}
\def\d{-- ++(0,-1)}
\def\ru{-- ++(0.8660254,0.5)}
\def\rd{-- ++(0.8660254,-0.5)}
\def\lu{-- ++(-0.8660254,0.5)}
\def\ld{-- ++(-0.8660254,-0.5)}

\def\ww{\r \u \r \u \r \d }
\def\wwa{\r \u \r \u \r \d }
\def\wwb{\u \r \u \r \d \r }
\def\wwc{\r \u \r \d \r \u }
\def\wwd{\u \r \d \r \u \r }
\def\wwe{\r \d \r \u \r \u }
\def\wwf{\d \r \u \r \u \r }

\def\nwa{\r \u \r \u \r \d }
\def\nwb{\u \l \l \l \u \u }
\def\nwc{\d \d \d \l \u \l }

\begin{document}

\title{Endless self-avoiding walks}
\author{Nathan Clisby \\
ARC Centre of Excellence for Mathematics and Statistics of
Complex Systems,\\
Department of Mathematics and Statistics,\\
The University of Melbourne, VIC 3010, Australia}
\date{March 13, 2013}
\maketitle

\begin{abstract}
We introduce a self-avoiding walk model for which end-effects are
completely eliminated. We enumerate the number of these walks 
for various lattices in dimensions two and three,
and use these enumerations to study the properties of this
model. 
We find that endless self-avoiding walks have the same connective constant as
self-avoiding walks, and the same Flory exponent $\nu$. However, there
is no power
law correction to the exponential number growth for this new model, i.e.
the critical exponent $\gamma = 1$ exactly in any dimension.
In addition, 
the number growth has no analytic corrections to scaling, and 
we have convincing numerical evidence to support the
conjecture that the amplitude for the number
growth is a universal quantity.
The technique by which
end-effects are eliminated may be generalized to other models of
polymers such as interacting self-avoiding walks.
\end{abstract}

\section{Introduction\label{sec:intro}}

Self-avoiding walks (SAW) on a regular lattice are an important model in
statistical mechanics with a long history~\cite{Madras1993}. An $n$-step
SAW is a map $\omega$ from the integers $\{0,1,\cdots,n\}$ to sites on
the lattice, with $\omega(0)$ conventionally at the origin, and
$|\omega(i+1) - \omega(i)| = 1$, and $\omega(i) \neq \omega(j) \forall i
\neq j$. The SAW model is the $N \rightarrow 0$ limit of the $N$-vector
model, which includes the Ising model as the $N = 1$ case, and as such
it is an important model for the theoretical understanding of critical
phenomena~\cite{deGennes1972}. Self-avoiding walks are in the same
universality class as polymers in a good solvent, and thus the model
also has an important physical interpretation. Much is known about
self-avoiding walks, particularly in dimensions $d \geq
4$~\cite{Hara1992,Brydges2010}, and considerable progress has been made
in understanding self-avoiding walks in two
dimensions~\cite{Nienhuis1982,Lawler2004,Duminil-Copin2012b}, but an
exact solution remains elusive for any dimension bar the trivial case
$d=1$.

The most important quantities which characterize SAW are the number of
SAW of length $n$, $c_n$, and measures of the size of the walk, such as
the square end-to-end distance, $r_n$. In the following definitions we assume
that summations are performed over the set of self-avoiding walks of
length $n$:
\begin{align}
c_n &= \sum_{|\omega|=n} 1 
\\ r_n &= \sum_{|\omega|=n} |\omega(n)|^2 
\end{align}
We abuse notation by allowing $|\omega|$ to measure the number of steps
of a walk $\omega$,
while $|\omega(n)|^2$ is the square of the
Euclidean norm of the site $\omega(n)$.

The asymptotic behavior of $c_n$ and $r_n$ is characterized by certain
quantities of fundamental interest, namely the connective constant,
$\mu$, which is lattice dependent, and the critical exponents $\gamma$
and $\nu$ which are universal. We have
\begin{align}
c_n &\sim A \mu^n n^{\gamma - 1} \left(1 +
\frac{a^*_1}{n} + \frac{a^*_2}{n^2} + \cdots + \frac{b^*_0}{n^{\Delta_1}} +
\cdots \right) + \text{AF}, \label{eq:cnasympt}
\\ r_n &\sim A D \mu^n n^{2\nu + \gamma - 1} \left(1 +
\frac{f^*_1}{n} + \frac{f^*_2}{n^2} + \cdots + \frac{g^*_0}{n^{\Delta_1}} +
\cdots \right) + \text{AF}, \label{eq:rnasympt}
\\ \rho_n & \sim D n^{2\nu} \left(1 +
\frac{l^*_1}{n} + \frac{l^*_2}{n^2} + \cdots + \frac{m^*_0}{n^{\Delta_1}} +
\cdots \right) + \text{AF}, \label{eq:rhonasympt}
\end{align}
where we define the mean square end-to-end distance as $\rho_n
\equiv r_n/c_n$.
The corrections in (\ref{eq:cnasympt})--(\ref{eq:rhonasympt}) with integer powers of $n$ are the analytic corrections
to scaling, while 
$\Delta_1$ is the exponent of the leading non-analytic correction to
scaling.
There are additional higher order analytic corrections with exponents
$-3, -4, \cdots$, and non-analytic corrections with exponents $\Delta_2,
\Delta_3, \cdots$. The higher order
corrections to scaling involve arbitrary mixed terms as well, of the
general form
$\text{const.}/n^{j_0+j_1\Delta_1+j_2\Delta+j_3\Delta_3+\cdots}$, with
non-negative integers $j_i$.
In (\ref{eq:cnasympt})--(\ref{eq:rhonasympt}), $\text{AF}$ denotes
the possible contribution of the so-called anti-ferromagnetic term which occurs for bipartite
lattices such as the square lattice and the simple cubic lattice. This
term alternates in sign, and is sub-leading as the exponent arises from
polygon-like configurations.

See
\cite{Caracciolo2005} for extensive discussions and analysis of the
correct asymptotic form for two-dimensional SAW, and \cite{Clisby2007}
for a discussion of the full asymptotic form for three-dimensional SAW.

Due to the importance of SAW, in the absence of an exact solution it is
desirable to find simpler models. Recent progress has been made in
finding interesting solvable models of walks, with non-trivial critical
exponents, such as prudent walks~\cite{Garoni2009,Bousquet-Melou2010a}.
Another recent paper~\cite{Gherardi2009,Gherardi2010} takes a different approach,
developing a prescription to discretize SLE, resulting in a
self-avoiding walk which is a finite piece of an infinite SLE curve.

Other models which are related to SAW are self-avoiding returns and
polygons (SAR and SAP respectively) and
bridges. A self-avoiding return (SAR) is a SAW which is forced to 
return to the origin, i.e. an $n$-step SAR is identified with an $n-1$-step SAW with
$|\omega(n-1)-\omega(0)| = 1$. An SAP is an \emph{unrooted}
\emph{unoriented} SAR, and so there are exactly $2n$ corresponding SAR
for each $n$-step SAP.
Bridges are a subset of SAW
with an additional condition on the $x$-coordinates of a walk: $\omega_1(0)
< \omega_1(i) \leq \omega_1(n)$ with $0 < i \leq n$.
Both SAP and bridges have the same connective constant as SAW, and the
same size exponent $\nu$ as SAW, but different critical exponents for
their growth in number.

In this paper we introduce a new variant of SAW, which we call endless
self-avoiding walks, or eSAW.  We will prove that this new model has the
same connective constant $\mu$ as SAW, and argue that it has the same
critical exponent $\nu$ as SAW but with number growth exponent
$\gammae=1$ in any dimension.
The number growth has no analytic corrections to
scaling, and we find strong numerical
evidence that the amplitude $\Ae$ of the number growth is a universal
quantity.

We will study eSAW on various two-dimensional and three-dimensional
lattices, in particular the square, honeycomb, triangular, union jack,
simple cubic, body centered cubic (bcc), and face centered cubic (fcc)
lattices.  The model can be naturally extended to other lattices, and
also to other models of walks as described in Sec.~\ref{sec:discussion}.

\subsection{Outline}

We will first introduce the endless self-avoiding walk model in
Sec.~\ref{sec:endless}, and discuss its expected properties. We will
then describe the enumeration method in Sec.~\ref{sec:method}, and
analyze the resulting enumerations to estimate critical parameters for
eSAW in Sec.~\ref{sec:analysis}. In Sec.~\ref{sec:discussion} we will
discuss various aspects of the eSAW model,
and finally conclude in Sec.~\ref{sec:conclusion}.

\section{Endless self-avoiding walks\label{sec:endless}}

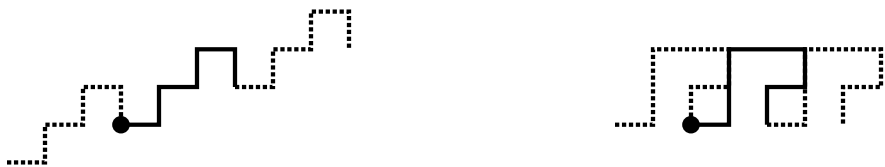
\begin{figure}[htb]
\begin{center}
\begin{tikzpicture}[ultra thick,scale=0.5]
\root
\sww \ww \ew
\draw[style=densely dotted] (3,1) \ww \ew
\draw[style=densely dotted] (-3,-1) \ww \ew
\begin{scope}[shift={(15,0)}]
\root
\sww \r \u \u \r \r \d \l \d \ew
\draw[style=densely dotted] (2,0) \r \u \u \r \r \d \l \d \ew
\draw[style=densely dotted] (-2,0) \r \u \u \r \r \d \l \d \ew
\end{scope}
\end{tikzpicture}
\end{center}
\caption{An example eSAW, $\wex$, on the left; on the right is a SAW which is not an
 eSAW. \label{fig:example}}
\end{figure}

The basic idea for constructing the model is extremely simple: we start
with the set of self-avoiding walks, and endless self-avoiding walks are
the subset of walks which are still self-avoiding after they have been
concatenated with copies of themselves head-to-tail \emph{ad infinitum}.
An example of an eSAW on the square lattice is shown on the left in
Fig.~\ref{fig:example}, with the root point shown as a solid circle and
the concatenated copies as dotted lines, while on the right is a SAW
which is not an eSAW. Because the walk is extended to infinity, sites on
the walk are indistinguishable and end-effects are completely
eliminated. This definition of eSAW is applicable for any lattice.

An alternative, illuminating definition of eSAW may be given in terms
of words, where we define the walk in terms of sequences of steps rather
than sites on a lattice. We now specialize to $\Z^d$ for the sake of
concreteness, but the argument is similar for other lattices. Let us
denote the steps in the positive axis directions as $1, 2, \cdots, d$,
and the steps in the negative axis directions as $\bar{1}, \bar{2},
\cdots, \bar{d}$. Then SAW of $n$ steps correspond to the set of words
which contain no loops, i.e. for which there are no sub-words with equal
numbers of $1$ and $\bar{1}$, $2$ and $\bar{2}$, $\cdots$, $d$ and
$\bar{d}$. Given a walk, the infinite word which results from
concatenating that word with itself indefinitely is an endless walk. If
the infinite word has no loops, then the walk is an endless SAW
in correspondence with our earlier definition. E.g., if we consider the
example SAW from Fig.~\ref{fig:example}, $\wex = 12121\bar{2}$, then the
corresponding infinite word is
$\cdots12121\bar{2}12121\bar{2}12121\bar{2}\cdots$. The infinite word has
no loops, and hence $\wex$ is an eSAW.

When we choose boundary conditions for a lattice model, conventionally
this boundary is in real space, i.e. it applies to a piece of the
lattice.  In contrast, we consider eSAW to be SAW with periodic
boundary conditions in what we could loosely describe as chain space. In
chain space, SAW and eSAW are one-dimensional objects, regardless of the
embedding dimension.

We are principally interested in the asymptotic behavior of the number
of eSAW of $n$ steps, $e_n$, and the square end-to-end distance $er_n$
(other quantities of interest are defined in Sec.~\ref{sec:sac}). In
analogy with (\ref{eq:cnasympt})--(\ref{eq:rhonasympt}) we expect
that:
\begin{align}
e_n &\sim \Ae \mue^n n^{\gammae - 1} \left(1 +
\frac{a_1}{n} + \frac{a_2}{n^2} + \cdots + \frac{b_0}{n^{\Delta_1}} +
\cdots \right) + \text{AF}, \label{eq:enasympt}
\\ er_n &\sim \Ae \De \mue^n n^{2\nue + \gammae - 1} \left(1 +
\frac{f_1}{n} + \frac{f_2}{n^2} + \cdots + \frac{g_0}{n^{\Delta_1}} +
\cdots \right) + \text{AF}, \label{eq:ernasympt}
\\ e\rho_n & \sim \De n^{2\nue} \left(1 +
\frac{l_1}{n} + \frac{l_2}{n^2} + \cdots + \frac{m_0}{n^{\Delta_1}} +
\cdots \right) + \text{AF}, \label{eq:erhonasympt}
\end{align}
where $e\rho_n \equiv er_n/e_n$.

In Sections~\ref{sec:proof}-\ref{sec:properties} we will specialize to
discussing eSAW on $\Z^d$, and will discuss the minor differences
arising for
the honeycomb lattice in Sec.~\ref{sec:honeycomb}.

\subsection{Proof that $\mu_{\rm e} = \mu$ for eSAW on $\Z^d$\label{sec:proof}}

Every bridge of $n$ steps is also an eSAW of $n$ steps, since by
construction bridges can be concatenated with themselves (indeed, any
bridge can be concatenated with any other bridge and remain
self-avoiding). It is also clear that every eSAW is in fact a SAW, and
hence if $\bridge_n$
is the number of bridges of $n$ steps,
$e_n$ is the number of eSAW of $n$ steps, and $c_n$ is the number of SAW
of $n$ steps, we must have
\begin{align}
\bridge_n &\leq e_n \leq c_n.
\end{align}
Therefore we must have
\begin{align}
\lim_{n \rightarrow \infty} \bridge_n^{1/n} &\leq 
 \lim_{n \rightarrow \infty} e_n^{1/n}
 \leq \lim_{n \rightarrow \infty} c_n^{1/n},
\end{align}
but it has been proved~\cite{Hammersley1957} that 
\begin{align}
\lim_{n \rightarrow \infty} \bridge_n^{1/n} &= 
 \lim_{n \rightarrow \infty} c_n^{1/n} = \mu
\end{align}
Thus, we have
\begin{align}
\mu_{\rm e} &\equiv \lim_{n \rightarrow \infty} e_n^{1/n} = \mu,
\end{align}
i.e. the connective constant for eSAW is precisely the same
as that for self-avoiding walks.

There is strong evidence that $\bridge_n$ and $c_n$ obey the scaling laws:
\begin{align}
\bridge_n &\sim n^{b-1} \mu^n
\\ c_n &\sim n^{\gamma - 1} \mu^n
\end{align}
where $b < 1$\footnote{In fact for two-dimensional lattices 
numerical evidence and SLE arguments provide convincing evidence
that $b = 9/16$. Private communication from A. J. Guttmann.}. This in turn implies that $b \leq \gammae \leq
\gamma$; we will argue in the following subsection that in fact
$\gammae = 1$.

\subsection{Argument that $\gammae = 1$\label{sec:gamma}}
We will now give a number of heuristic arguments in support of our
expectation that $\gammae = 1$.
Our intent is to communicate the intuition we have as to \emph{why}
$\gammae = 1$.

Direct renormalization group arguments regarding the partition function
for linear polymer models such as SAW in two and three dimensions lead
to $\gamma > 1$~\cite{deGennes1979,Duplantier1989,Oono1981}.
Heuristically, this is due to the fact that ends have more freedom than
the interior of the chain, and the degree of freedom increases as the
length of the chain increases. See Chapter XI.1.6 of \cite{deGennes1979}
for a qualitative but more detailed argument, and \cite{Duplantier1989} for
theoretical arguments regarding polymer networks with an arbitrary
number of ends. See \cite{Oono1981} for a systematic renormalization
group calculation for linear polymers in conformation space, which
explicitly showed the leading order contribution which perturbs $\gamma$
away from 1 is due
to an end effect. 

For clarity we reproduce part of the argument of \cite{deGennes1979}
here.  Suppose we can express the free energy $F$ of a polymer chain in
terms of length $n$ and interaction strength $u$. Then, if we
renormalize the chain by a scale factor $g$ by integrating out the
structure at small length scales, we would expect that we could express
the free energy of the full chain in terms of the free energy of the
renormalized chain as follows:
\begin{align}
F(n,u) &= \Delta F(n,u) + F(n/g,u_1).\label{eq:freeenergy}
\end{align}
We are interested in the behavior of $F$ in the vicinity
of the fixed point $u^*$, where $u \approx u_1 \approx u^*$.
For SAW, $\Delta F$ includes a contribution from interior chain segments,
which have two neighbors, and 
an additional contribution from the ends, which only interact with one neighbor. It
is the contribution of the ends that perturbs $\gamma$ away from 1.

For SAW, the enhancement effect becomes negligible for dimensions
$d > 4$ where $\gamma=1$, while there is a logarithmic correction to
the number growth in dimension four. For models where the ends
have less freedom than the interior of the chain, such as bridges and
SAP, we would expect (and do indeed observe) that the growth exponent is
strictly less than unity.

Now, the complete absence of any end-effects for eSAW should mean
that $\gamma$ is \emph{precisely} equal to 1, as there is no physical
mechanism to perturb the value of $\gamma$.
For eSAW, $\Delta F$ in (\ref{eq:freeenergy}) only includes contributions from interior chain
segments, and hence $\gammae = 1$.

We may consider eSAW as an object extending to infinity (a rooted
self-avoiding chain, in the notation of Sec.~\ref{sec:sac}). When we
directly renormalize a configuration, its global nature remains
invariant as it is still extends to infinity, and only the local
structure of the walk changes. The one large scale alteration is the
size of the repeating unit, or alternatively the density of boundaries along
the walk. For SAW, the sizes of the end-regions of the walks scale in a
different way to the interior of the chain.
That is, for SAW there
is an internal length scale (distance from the ends) which must be taken
into account when the system is renormalized, but for eSAW there is no
such length scale to worry about.

The ability to derive a model with $\gamma = 1$ via the imposition of
periodic boundary conditions is thus seen as
fundamentally due to the fact that linear polymers are one-dimensional
objects, no matter the dimensionality of the space they are embedded in. We
therefore believe it is unlikely that other lattice models can be
adapted in the same way. Of course, this does not preclude the
possibility that a new insight could lead to non-trivial
two-dimensional and three-dimensional models with $\gamma = 1$.
 
We note in passing that the process by which $\gamma$ may be altered
from its standard value is somewhat subtle: theoretical arguments
suggested that it should be possible to vary $\gamma$~\cite{Cardy1994}
by introducing orientation dependent interactions into the polymer
model. In particular, it was predicted that $\gamma$ would be different
for SAW on the Manhattan lattice~\cite{Bennett-Wood1998a} than on other
lattices due to the fact that SAW cannot be trapped on this lattice
(effectively introducing an orientation dependent interaction). However,
detailed numerical investigation~\cite{Caracciolo1999} subsequently
showed that $\gamma$ was unchanged from its value for SAW on other
two-dimensional lattices.

Another `endless' walk model is of course the model of self-avoiding
returns. The number of self-avoiding returns grows as $n^{\alpha-2}
\mu^n$, and the hyperscaling relation $d\nu = 2-\alpha$ can be
interpreted in the following way. SAW with free end-points have typical
size $n^\nu$, and so the fraction which return to the origin should
be $O(n^{-d\nu})$. Thus we should expect the number of SAR to be
\begin{align}
\#(SAR) &\sim n^a n^{-d\nu} \mu^n,
\end{align}
where $n^{-d\nu}$ is the factor due to geometric restriction, and $n^a$
is an enhancement factor which parallels the $n^{\gammae -1}$
factor for eSAW. 
There is strong theoretical and numerical evidence in support
of the hyperscaling relation, which implies that $a = 0$. Since there
are no geometric restrictions for eSAW, we regard the validity of the
hyperscaling relation for self-avoiding returns to provide support for
our argument that $\gammae=1$.

We readily acknowledge that our arguments for $\gammae = 1$ fall far
short of a proof, and will verify $\gammae = 1$ to good precision
via series analysis methods in Sec.~\ref{sec:analysis}.

\subsection{Argument that $\nue = \nu$\label{sec:nu}}

eSAW are different from SAW in that they have no ends, and so in a small
region of the chain near the end one may expect that, on average, eSAW
and SAW are qualitatively different. However, the size of SAW and
similar objects is
dominated by the region of the chain far from the end, and this is why
SAW, SAP, bridges, and eSAW all have the same size exponent $\nu$.

More explicitly, following Caracciolo et
al.~\cite{Caracciolo2000a}, we define
\begin{align}
\xi^2_n &= \frac{1}{2d} \rho_n,
\\ \varrho &= \frac{|\mathbf{r}|}{\xi_n}.
\end{align}
Then in the limit $n \rightarrow \infty$, $|\mathbf{r}| \rightarrow
\infty$, $\varrho$ fixed,
we expect that the end-to-end distribution function for
SAW is~\cite{Fisher1966,Caracciolo2000a}
\begin{align}
P(\varrho) &= \frac{1}{\xi_n^d} f(\varrho) \left(1 +
O(n^{-\Delta_1})\right)
\end{align}
where for large $\varrho$
\begin{align}
f(\varrho) &\approx f_\infty \varrho^\sigma \exp(-D\varrho^\delta),
\label{eq:enddistribution}
\end{align}
with $\sigma$ and $\delta$ both positive.
Now, eSAW are a subset of SAW, and so
if we suppose that $\nue > \nu$, then 
(\ref{eq:enddistribution}) would imply that $e_n$ could not
grow asymptotically as $e_n \sim n^a \mu^n$.
The most reasonable interpretation, based on the assumption that
$e_n \sim n^a \mu^n$ and $e\rho_n \sim n^{2\nue}$, is that $\nue \le
\nu$.

In addition, we expect that eSAW which return close to the origin are
suppressed, i.e. polygon-like configurations are relatively rarer
amongst eSAW than they are amongst SAW. The endless condition induces an
effective repulsion between the end-points, and thus we would expect that
$\nue \ge \nu$.
Therefore we conclude that $\nue = \nu$.

As an aside, we expect that the end-point repulsion would result in a
larger amplitude for the mean square end-to-end distance for eSAW than
for SAW.

\subsection{Analytic corrections to scaling\label{sec:analytic}}

For SAP on two-dimensional lattices, there is strong numerical evidence
that the
\emph{non-analytic}
corrections to scaling fold into the analytic background
term~\cite{Conway1996,Jensen1999,Caracciolo2005}.

For eSAW, we expect that the \emph{analytic} corrections to scaling will
fold into the analytic background term, due to the fact that $\gammae =
1$. If we define the generating function for the number of eSAW as
\begin{align}
\Ce(x) &= \sum_{n=1}^{\infty} e_n x^n,
\end{align}
then we expect that the asymptotic behavior of the $e_n$ series is
determined by the singularity or singularities of $\Ce(x)$ which are
closest to the origin. Neglecting the sub-leading anti-ferromagnetic
singularity (which may have zero amplitude for eSAW in any case; see
Sec.~\ref{sec:af}), we expect that 
\begin{align}
\Ce(x) &= \mathcal{A}(x) + \mathcal{B}(x) (1 - x/x_c)^{-\gammae} \left(1
+ c \, (1 - x/x_c)^{\Delta_1} + \cdots\right),
\end{align}
where $\mathcal{A}(x)$ and $\mathcal{B}(x)$ are analytic functions. If
we write
\begin{align}
\mathcal{B}(x) &= \Ae + \beta_1 (1-x/x_c) + \beta_2 (1-x/x_c)^2 +
 \beta_3 (1-x/x_c)^3 + \cdots,
\end{align}
then, under the assumption that $\gammae = 1$, all of the higher order
terms with coefficients $\beta_1, \beta_2, \beta_3,\cdots$ will be
absorbed into the analytic background term $\mathcal{A}(x)$!

Thus, in (\ref{eq:enasympt}) we expect that $a_1 = a_2 = \cdots
= 0$.  We will test this hypothesis, as there could conceivably be a
confluent logarithmic correction which would result in non-zero
coefficients $a_i$.

\subsection{Irreducible eSAW and self-avoiding chains \label{sec:sac}}

The more natural definition of an endless SAW (to us), which we will
call a self-avoiding chain (SAC), is related to
eSAW in the same way that self-avoiding polygons are related to
self-avoiding returns. SAC are equivalence classes of
eSAW which correspond to the same `shape', regardless of the root point,
 or the direction of traversal.

\begin{figure}[htb]
\begin{center}
\begin{tikzpicture}[ultra thick,scale=0.5]
\begin{scope}[shift={(0,0)}]
\draw[fill] (0,0) circle (5pt);
\sww 
\r \u \r \u \r \d \ew
\draw[style=densely dotted] (3,1) 
\ww \ww \ew
\draw[style=densely dotted] (-6,-2) 
\ww \ww \ew
\end{scope}
\begin{scope}[shift={(9,-2)}]
\draw[fill] (0,0) circle (5pt);
\sww 
\r \u \r \u \r \d \r \u \r \u \r \d \ew
\draw[style=densely dotted] (6,2) 
\r \u \r \u \r \d \r \u \r \u \r \d \ew
\draw[style=densely dotted] (-6,-2) 
\r \u \r \u \r \d \r \u \r \u \r \d \ew
\end{scope}
\end{tikzpicture}
\end{center}
\caption{Examples of two eSAW which correspond to the same infinite
 chain. The eSAW on the left is irreducible.\label{fig:irreducible}}
\end{figure}
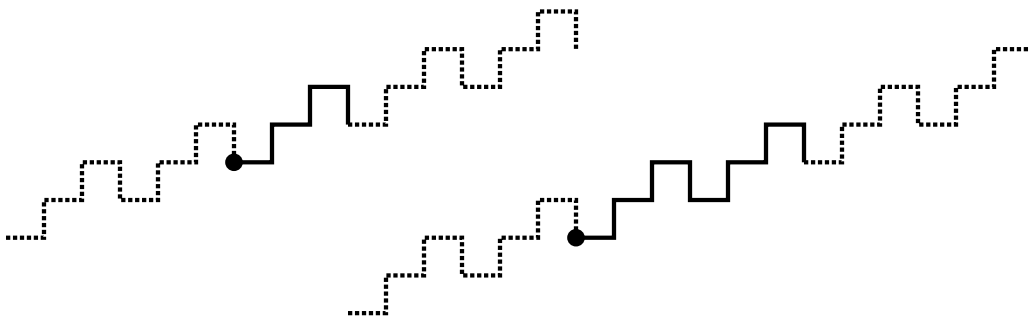

For an $n$-step eSAW on $\Z^d$ there are $n$ possible root points,
and two directions of traversal, suggesting that the number of eSAW
of $n$ steps
should be $2 n$ times the number of SAC.
However, there is an additional complication
for the eSAW / SAC correspondence which does not occur for
self-avoiding returns / self-avoiding polygons: eSAW can have
repeating sub-units which makes the infinite chain corresponding to
a given eSAW non-unique. 
We illustrate this in
Fig.~\ref{fig:irreducible}, where it can be seen that the same
infinite chain results for the two eSAW shown. We call those eSAW
with no repeating units \emph{irreducible} eSAW. The number of
eSAW, $e_n$, and the square end-to-end distance $r_n$
can then be straightforwardly expressed in terms of the number
irreducible eSAW, $i_n$, and the square end-to-end distance of
irreducible eSAW, $ir_n$:
\begin{align}
e_n &= \sum_{j | n} i_j, \\
er_n &= \sum_{j | n} \frac{n^2}{j^2} ir_j.
\end{align}
The inverse expressions can be obtained, if desired, via a M\"obius
transform involving the M\"obius function $\mu$:
\begin{align}
i_n &= \sum_{j | n} \mu\left(\frac{n}{j}\right) e_j,
\\ ir_n &= \sum_{j | n} \mu\left(\frac{n}{j}\right) \frac{n^2}{j^2}
er_j.
\end{align}
An alternative interpretation is that the word corresponding to an
irreducible eSAW cannot be decomposed into identical, repeating
sub-words.

Note that one does not need to worry about ``irreducible'' SAR
configurations precisely because of the self-avoidance condition. If we
were to consider \emph{weakly} as opposed to \emph{strictly} avoiding
self-avoiding returns / polygons instead then the same complication
would occur, as loops could then repeat.

\begin{figure}[htb]
\begin{center}
\begin{tikzpicture}[ultra thick,scale=0.5]
\begin{scope}[shift={(0,0)}]
\draw[fill] (0,0) circle (5pt);
\sww \wwa \ew
\begin{scope}[shift={(3,1)}]
\sdw \wwa \ew
\end{scope}
\begin{scope}[shift={(-3,-1)}]
\sdw \wwa \ew
\end{scope}
\end{scope}
\begin{scope}[shift={(0,-4)}]
\draw[fill] (0,0) circle (5pt);
\sww \wwb \ew
\begin{scope}[shift={(3,1)}]
\sdw \wwb \ew
\end{scope}
\begin{scope}[shift={(-3,-1)}]
\sdw \wwb \ew
\end{scope}
\end{scope}
\begin{scope}[shift={(0,-8)}]
\draw[fill] (0,0) circle (5pt);
\sww \wwc \ew
\begin{scope}[shift={(3,1)}]
\sdw \wwc \ew
\end{scope}
\begin{scope}[shift={(-3,-1)}]
\sdw \wwc \ew
\end{scope}
\end{scope}
\begin{scope}[shift={(12,0)}]
\draw[fill] (0,0) circle (5pt);
\sww \wwd \ew
\begin{scope}[shift={(3,1)}]
\sdw \wwd \ew
\end{scope}
\begin{scope}[shift={(-3,-1)}]
\sdw \wwd \ew
\end{scope}
\end{scope}
\begin{scope}[shift={(12,-4)}]
\draw[fill] (0,0) circle (5pt);
\sww \wwe \ew
\begin{scope}[shift={(3,1)}]
\sdw \wwe \ew
\end{scope}
\begin{scope}[shift={(-3,-1)}]
\sdw \wwe \ew
\end{scope}
\end{scope}
\begin{scope}[shift={(12,-8)}]
\draw[fill] (0,0) circle (5pt);
\sww \wwf \ew
\begin{scope}[shift={(3,1)}]
\sdw \wwf \ew
\end{scope}
\begin{scope}[shift={(-3,-1)}]
\sdw \wwf \ew
\end{scope}
\end{scope}
\end{tikzpicture}
\end{center}
\caption{Irreducible eSAW which are equivalent up to the root point.\label{fig:roots}}
\end{figure}
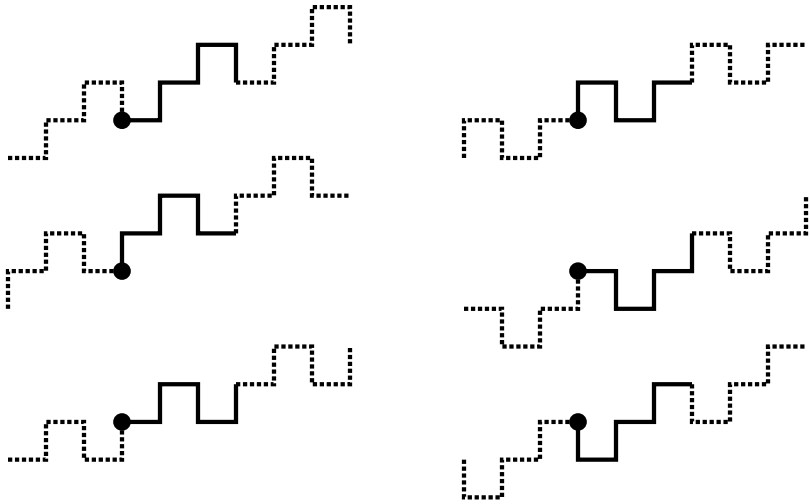

Now for each irreducible eSAW of $n$ steps there are $n$ possible root
points, as can be seen for our example eSAW in Fig.~\ref{fig:roots}. In
addition there are two possible orientations for the walk, and hence
there $2 n$ irreducible eSAW for each self-avoiding chain. If we
denote the number of SAC with an irreducible unit of $n$ steps as $d_n$,
then
\begin{align}
i_n &= 2 n d_n.
\end{align}

In terms of words, converting from rooted to unrooted eSAW can be done
by enforcing an ordering on the set of words corresponding to a given
chain, and then choosing a single representative of this set. A natural
way to do this is to use lexicographic ordering by placing an order on
the steps themselves, e.g. $1 < \bar{1} < 2 < \bar{2}$. 
In Fig.~\ref{fig:roots} this results in the ordering $12121\bar{2} <
121\bar{2}12 < 1\bar{2}1212 < 2121\bar{2}1 < 21\bar{2}121<
\bar{2}12121$; the eSAW in the top left of the figure, $12121\bar{2}$, 
would thence be chosen as the representative.
Orientation independence
on words can be enforced by defining words which are mapped to each
other under the involution $1 \leftrightarrow \bar{1}$ and $2
\leftrightarrow \bar{2}$ to be equivalent.

In summary, endless walks may be
reducible or irreducible,
rooted or unrooted,
and oriented or unoriented.
Using this taxonomy, the objects we will study in this paper are shown
in Table~\ref{tab:def}.
We expect that the asymptotic behavior of the irreducible and reducible
series will be \emph{exactly} the same, i.e. $i_n \sim e_n$ and $ir_n
\sim er_n$. Since $d_n = i_n/(2 n)$, the exponent for the growth in
$d_n$ is $(\gammae - 2)$ instead of $(\gammae - 1)$, but
otherwise the asymptotic growth for $d_n$ is given in 
 (\ref{eq:enasympt}).

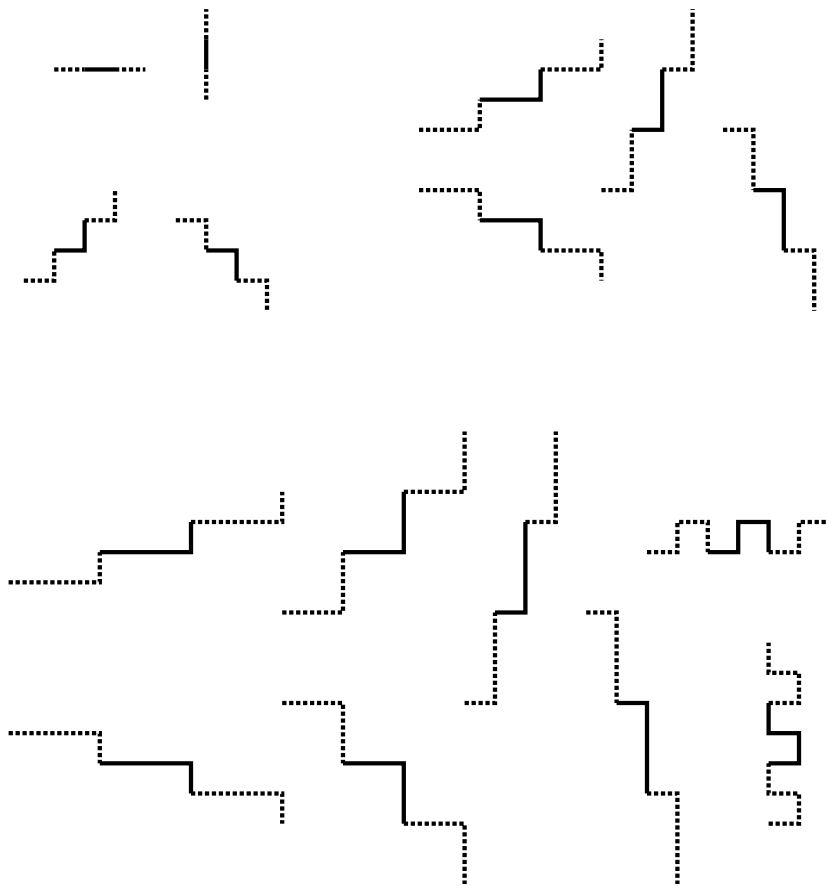
\begin{figure}[htb]
\begin{center}
\begin{tikzpicture}[ultra thick,scale=0.4]
\begin{scope}[shift={(0,0)}]
\begin{scope}[shift={(1,0)}]
\sww \r \ew
\begin{scope}[shift={(-1,0)}]
\sdw \r \ew
\end{scope}
\begin{scope}[shift={(1,0)}]
\sdw \r \ew
\end{scope}
\end{scope}
\begin{scope}[shift={(5,0)}]
\sww \u \ew
\begin{scope}[shift={(0,-1)}]
\sdw \u \ew
\end{scope}
\begin{scope}[shift={(0,1)}]
\sdw \u \ew
\end{scope}
\end{scope}
\end{scope}
\begin{scope}[shift={(0,-6)}]
\begin{scope}[shift={(0,0)}]
\sww \r \u \ew
\begin{scope}[shift={(-1,-1)}]
\sdw \r \u \ew
\end{scope}
\begin{scope}[shift={(1,1)}]
\sdw \r \u \ew
\end{scope}
\end{scope}
\begin{scope}[shift={(5,0)}]
\sww \r \d \ew
\begin{scope}[shift={(-1,1)}]
\sdw \r \d \ew
\end{scope}
\begin{scope}[shift={(1,-1)}]
\sdw \r \d \ew
\end{scope}
\end{scope}
\end{scope}
\begin{scope}[shift={(14,-1)}]
\begin{scope}[shift={(0,0)}]
\sww \r \r \u \ew
\begin{scope}[shift={(-2,-1)}]
\sdw \r \r \u \ew
\end{scope}
\begin{scope}[shift={(2,1)}]
\sdw \r \r \u \ew
\end{scope}
\end{scope}
\begin{scope}[shift={(0,-4)}]
\sww \r \r \d \ew
\begin{scope}[shift={(-2,1)}]
\sdw \r \r \d \ew
\end{scope}
\begin{scope}[shift={(2,-1)}]
\sdw \r \r \d \ew
\end{scope}
\end{scope}
\begin{scope}[shift={(5,-1)}]
\sww \r \u \u \ew
\begin{scope}[shift={(-1,-2)}]
\sdw \r \u \u \ew
\end{scope}
\begin{scope}[shift={(1,2)}]
\sdw \r \u \u \ew
\end{scope}
\end{scope}
\begin{scope}[shift={(9,-3)}]
\sww \r \d \d \ew
\begin{scope}[shift={(-1,2)}]
\sdw \r \d \d \ew
\end{scope}
\begin{scope}[shift={(1,-2)}]
\sdw \r \d \d \ew
\end{scope}
\end{scope}
\end{scope}
\begin{scope}[shift={(1.5,-16)}]
\begin{scope}[shift={(0,0)}]
\sww \r \r \r \u \ew
\begin{scope}[shift={(-3,-1)}]
\sdw \r \r \r \u \ew
\end{scope}
\begin{scope}[shift={(3,1)}]
\sdw \r \r \r \u \ew
\end{scope}
\end{scope}
\begin{scope}[shift={(0,-7)}]
\sww \r \r \r \d \ew
\begin{scope}[shift={(-3,1)}]
\sdw \r \r \r \d \ew
\end{scope}
\begin{scope}[shift={(3,-1)}]
\sdw \r \r \r \d \ew
\end{scope}
\end{scope}
\begin{scope}[shift={(8,0)}]
\sww \r \r \u \u \ew
\begin{scope}[shift={(-2,-2)}]
\sdw \r \r \u \u \ew
\end{scope}
\begin{scope}[shift={(2,2)}]
\sdw \r \r \u \u \ew
\end{scope}
\end{scope}
\begin{scope}[shift={(8,-7)}]
\sww \r \r \d \d \ew
\begin{scope}[shift={(-2,2)}]
\sdw \r \r \d \d \ew
\end{scope}
\begin{scope}[shift={(2,-2)}]
\sdw \r \r \d \d \ew
\end{scope}
\end{scope}
\begin{scope}[shift={(13,-2)}]
\sww \r \u \u \u \ew
\begin{scope}[shift={(-1,-3)}]
\sdw \r \u \u \u \ew
\end{scope}
\begin{scope}[shift={(1,3)}]
\sdw \r \u \u \u \ew
\end{scope}
\end{scope}
\begin{scope}[shift={(17,-5)}]
\sww \r \d \d \d \ew
\begin{scope}[shift={(-1,3)}]
\sdw \r \d \d \d \ew
\end{scope}
\begin{scope}[shift={(1,-3)}]
\sdw \r \d \d \d \ew
\end{scope}
\end{scope}
\begin{scope}[shift={(20,0)}]
\sww \r \u \r \d \ew
\begin{scope}[shift={(-2,0)}]
\sdw \r \u \r \d \ew
\end{scope}
\begin{scope}[shift={(2,0)}]
\sdw \r \u \r \d \ew
\end{scope}
\end{scope}
\begin{scope}[shift={(22,-7)}]
\sww \r \u \l \u \ew
\begin{scope}[shift={(0,-2)}]
\sdw \r \u \l \u \ew
\end{scope}
\begin{scope}[shift={(0,2)}]
\sdw \r \u \l \u \ew
\end{scope}
\end{scope}
\end{scope}
\end{tikzpicture}
\end{center}
\caption{The self-avoiding chains with irreducible units of 1-4 steps on
the square lattice. By inspection, $d_1=2$, $d_2=2$, $d_3=4$, and $d_4=8$, in
agreement with the values in Table~\ref{tab:z2} of
\ref{sec:data}.\label{fig:sac}}
\end{figure}

We show the self-avoiding chains with irreducible units of 1-4 steps
in Fig.~\ref{fig:sac}.

\begin{table}[ht]
\begin{center}
\begin{tabular}{cll}
\hline
Object & Properties & Sequence \\
\hline
\multirow{3}{*}{eSAW} & reducible, rooted, oriented &$e_n$ \\
& square end-to-end distance of eSAW & $er_n$ \\
& mean square end-to-end distance of eSAW & $e\rho_n$ \\[1ex]
\multirow{3}{*}{irreducible eSAW} & irreducible, rooted, oriented&$i_n$\\
& square end-to-end distance & $ir_n$ \\
& mean square end-to-end distance & $i\rho_n$ \\[1ex]
\multirow{1}{*}{self-avoiding chains} & irreducible, unrooted, unoriented & $d_n$\\
\hline
\end{tabular}
\caption{Definitions of objects and their corresponding sequences.\label{tab:def}}
\end{center}
\end{table}

\subsection{Properties of eSAW\label{sec:properties}}

We will now briefly mention some properties of eSAW which may not be
immediately apparent.

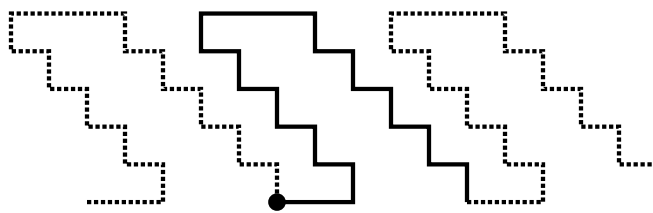
\begin{figure}[htb]
\begin{center}
\begin{tikzpicture}[ultra thick,scale=0.5]
\begin{scope}[shift={(0,0)}]
\draw[fill] (0,0) circle (5pt);
\sww 
\r \r \u \l \u \l \u \l \u \l \u \r \r \r \d \r \d \r \d \r \d \r \d
\ew
\begin{scope}[shift={(-5,0)}]
\sdw
\r \r \u \l \u \l \u \l \u \l \u \r \r \r \d \r \d \r \d \r \d \r \d
\ew
\end{scope}
\begin{scope}[shift={(5,0)}]
\sdw
\r \r \u \l \u \l \u \l \u \l \u \r \r \r \d \r \d \r \d \r \d \r \d
\ew
\end{scope}
\end{scope}
\end{tikzpicture}
\end{center}
\caption{An eSAW for which no cyclic permutation is a bridge.\label{fig:notbridge}}
\end{figure}

Firstly, although all bridges are eSAW, the converse is not true. For
example, the eSAW in Fig.~\ref{fig:notbridge} is not a bridge, and no
choice of start and end points on the self-avoiding chain corresponding
to the eSAW results in a bridge. A necessary and sufficient condition
for a SAC to allow for the choice of a bridge as the repeating unit, or
equivalently for the word representation of an eSAW to have a cyclic
permutation which is a bridge, is that there must exist a vertical line
which cuts only a single bond of the SAC. The proof of this statement is
obvious by inspection.

Going further, a natural generalization of bridges would be to allow for
directions other than vertical for the cutting line. If all SAC were
to admit at least one straight line which cuts only a single bond,
then we would regard eSAW as a generalization of bridges. However, it
is straightforward to derive eSAW for which any straight line must past
through at least two bonds of the SAC, for example by
slightly perturbing the eSAW in Fig.~\ref{fig:notbridge}.

\begin{figure}[htb]
\begin{center}
\begin{tikzpicture}[ultra thick,scale=0.5]
\begin{scope}[shift={(0,0)}]
\draw[fill] (0,0) circle (5pt);
\sww 
\r \d \r \d \r \r \r \r \r \r \u \r \u \r \u \u \u \l \u \l \l \l \l \d \l \d \d \d \d
\r \u \r \u \r
\ew
\end{scope}
\begin{scope}[shift={(14,0)}]
\draw[fill] (0,0) circle (5pt);
\sww 
\r \d \r \d \r \r \r \r \r \r \u \r \u \r \u \u \u \l \u \l \l \l \l \d \l \d \d \d
\r \u \r \u \r
\ew
\begin{scope}[shift={(7,2)}]
\sdw 
\r \d \d \l \d \l \l \l \l \u \l \u \u
\ew
\end{scope}
\end{scope}
\end{tikzpicture}
\end{center}
\caption{In solid lines, a trapped SAW on the left and a non-trapped SAW on the right
(a possible continuation is shown as a dotted line).
Neither of these are eSAW, but the non-trapped SAW may be a sub-walk of
an eSAW.\label{fig:trapped}}
\end{figure}
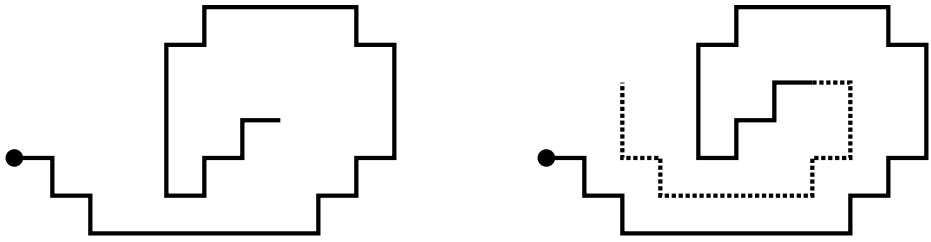

Trapped self-avoiding walks are SAW which cannot be extended
indefinitely from (at least) one of the ends and still remain
self-avoiding (see e.g. \cite{Chan2012}). 
By definition, no trapped self-avoiding walk is an eSAW. Conversely,
every non-trapped self-avoiding walk is a sub-walk of an infinite number
of eSAW, since the walk can always be extended far away from any
problematic regions.
Equivalently, all eSAW are Kesten
patterns~\cite{Kesten1963}, and we expect that all patterns have
positive density in sufficiently long eSAW.
Examples of trapped and non-trapped SAW are shown in
Fig.~\ref{fig:trapped}. 

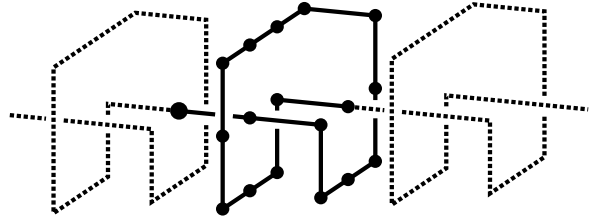
\begin{figure}[htb]
\begin{center}
\tdplotsetmaincoords{-75}{21}
\begin{tikzpicture}[scale=1.0,ultra thick,tdplot_main_coords]
\draw[fill] (0,0,0) circle (2.5pt);
\draw[style=densely dotted] (-2,-1,0) 
-- ++(0.51,0,0) ++(0.25,0,0) -- ++(1.24,0,0)
-- ++(0,0,1)
-- ++(0,2,0)
-- ++(0,0,-0.59) ++(0,0,-0.25) -- ++(0,0,-1.16)
-- ++(-1,0,0)
-- ++(0,-3,0)
-- ++(0,0,2)
-- ++(0,2,0)
-- ++ (0,0,-0.6) ++(0,0,-0.25) -- ++(0,0,-0.15)
-- ++(1,0,0);
\draw (0,0,0) 
-- ++(0.51,0,0) ++(0.25,0,0) -- ++(1.24,0,0)
-- ++(0,0,1)
-- ++(0,2,0)
-- ++(0,0,-0.59) ++(0,0,-0.25) -- ++(0,0,-1.16)
-- ++(-1,0,0)
-- ++(0,-3,0)
-- ++(0,0,2)
-- ++(0,2,0)
-- ++ (0,0,-0.6) ++(0,0,-0.25) -- ++(0,0,-0.15)
-- ++(1,0,0);
\path (0,0,0) 
++(1,0,0) [fill] circle(2.5pt)
++(1,0,0) [fill] circle(2.5pt)
++(0,0,1) [fill] circle(2.5pt)
++(0,1,0) [fill] circle(2.5pt)
++(0,1,0) [fill] circle(2.5pt)
++(0,0,-1) [fill] circle(2.5pt)
++(0,0,-1) [fill] circle(2.5pt)
++(-1,0,0) [fill] circle(2.5pt)
++(0,-1,0) [fill] circle(2.5pt)
++(0,-1,0) [fill] circle(2.5pt)
++(0,-1,0) [fill] circle(2.5pt)
++(0,0,1) [fill] circle(2.5pt)
++(0,0,1) [fill] circle(2.5pt)
++(0,1,0) [fill] circle(2.5pt)
++(0,1,0) [fill] circle(2.5pt)
++(0,0,-1) [fill] circle(2.5pt)
++(1,0,0) [fill] circle(2.5pt);
\draw[style=densely dotted] (2,1,0) 
-- ++(0.51,0,0) ++(0.25,0,0) -- ++(1.24,0,0)
-- ++(0,0,1)
-- ++(0,2,0)
-- ++(0,0,-0.59) ++(0,0,-0.25) -- ++(0,0,-1.16)
-- ++(-1,0,0)
-- ++(0,-3,0)
-- ++(0,0,2)
-- ++(0,2,0)
-- ++ (0,0,-0.6) ++(0,0,-0.25) -- ++(0,0,-0.15)
-- ++(2,0,0);
\end{tikzpicture}
\end{center}
\caption{A knotted eSAW.\label{fig:knot}}
\end{figure}

Unlike SAW, eSAW in dimensions $d > 2$ allow for true knots to occur
when we consider them as infinite chains.  An example knotted eSAW is
shown in Fig.~\ref{fig:knot}; it corresponds to the word
$11\bar{2}\bar{3}\bar{3}22\bar{1}333\bar{2}\bar{2}\bar{3}\bar{3}21$, and
is of length 17. This compares to the shortest knotted SAP on the simple
cubic lattice, which has 24 steps~\cite{Diao1993,Hong2013}. We do not
know if 17 is the minimum length for an eSAW knot on the simple cubic
lattice.

Finally we note that in two-dimensions all eSAW must have zero winding
angle. The winding angle has featured in recent
proofs of exact results for self-avoiding
walks~\cite{Duminil-Copin2012b,Beaton2012c}.

\subsection{eSAW on the honeycomb lattice\label{sec:honeycomb}}

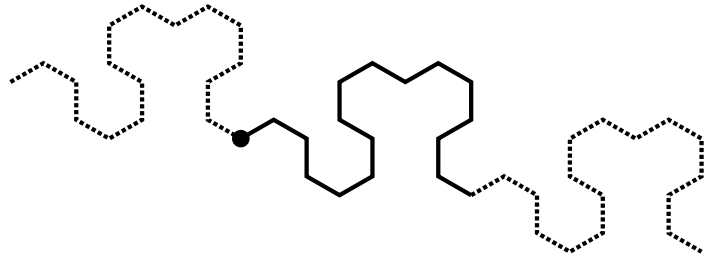
\begin{figure}[htb]
\begin{center}
\begin{tikzpicture}[ultra thick,scale=0.5]
\root
\sww \ru \rd \d \rd \ru \u \lu \u \ru \rd \ru \rd \d \ld \d \rd \ew
\draw[style=densely dotted] (-6.0621778265,1.5) \ru \rd \d \rd \ru \u \lu \u \ru \rd \ru \rd \d \ld \d \rd \ew
\draw[style=densely dotted] (6.0621778265,-1.5) \ru \rd \d \rd \ru \u \lu \u \ru \rd \ru \rd \d \ld \d \rd \ew
\end{tikzpicture}
\end{center}
\caption{Example eSAW on the honeycomb lattice. Note that the repeating
unit has an even number of steps; if one step were removed then the
eSAW would no longer be able to remain on the honeycomb lattice.\label{fig:honeycombexample}}
\end{figure}

The set of direction vectors for the neighboring sites of the honeycomb
lattice depends upon which of two bipartite sub-lattices one is
currently on. This contrasts with the other lattices considered in this
paper which possess full translation invariance. As a consequence, our
definition of eSAW implies that on the honeycomb lattice eSAW may only
have an even number of steps. 
An example of an eSAW for the honeycomb
lattice is shown in Fig.~\ref{fig:honeycombexample}. 
This in turn means that a $2n$ step
self-avoiding chain on the honeycomb lattice has $n$ possible root
points, rather than $2n$. The relations
between the different sequences must be modified accordingly:
\begin{align}
e_{2n} &= \sum_{j | n} i_{2j},
\\ er_{2n} &= \sum_{j | n} \frac{n^2}{j^2} ir_{2j},
\\ i_{2n} &= \sum_{j | n} \mu\left(\frac{n}{j}\right) e_{2j},
\\ ir_{2n} &= \sum_{j | n} \mu\left(\frac{n}{j}\right) \frac{n^2}{j^2}
er_{2j},
\\ i_{2n} &= 2 n d_{2n},
\end{align}
where once again $\mu$ is the M\"obius function.

It is possible to broaden the definition of eSAW to allow for eSAW with
an odd number of steps by altering the boundary conditions, e.g. by
interchanging left turns and right turns at the concatenation point.
However, this introduces some technical difficulties\footnote{E.g., in
this context it seems more natural to use `mid edges' as in
\cite{Duminil-Copin2012b} to define an eSAW with twisted boundary
conditions, because a single edge does
not carry any information about whether a left or right turn comes next.
}, and in the absence of a pressing reason to do so we will not explore
this option further here.

One key advantage for the analysis of eSAW on the honeycomb lattice
is that the connective constant is known to be exactly
$\sqrt{2+\sqrt{2}}$~\cite{Nienhuis1982,Duminil-Copin2012b}. We will exploit this fact
later in our analysis in Sec.~\ref{sec:analysis}.

The arguments for $\nue$ and $\gammae$ are not
altered for the honeycomb lattice.
However, the proof that $\mu_{\rm e} =
\mu$ would need to be modified. We will not prove that $\mu_{\rm e} =
\mu$ here, but are confident that this statement is indeed correct.

\section{Enumeration method\label{sec:method}}

We performed enumerations on the square, honeycomb, union
jack, simple cubic, bcc, and fcc lattices
using a basic backtracking algorithm. Backtracking works by explicitly
generating all of the combinatorial objects of interest by recursively
building up these objects from smaller objects. 

We will not describe backtracking in detail, but note that it has been
used to study SAW and related models for over 50 years (see section 7.3
of \cite{Hughes1995} for references). Here we focus on the differences
between our algorithm for enumerating eSAW and the standard backtracking
approach for enumerating SAW. More details on how to implement a
backtracking algorithm for the enumeration of SAW are given
in~\cite{Clisby2007}. 

The key difference between backtracking enumeration of eSAW and SAW, is
that as an eSAW is built up step by step one has to check for
self-intersections for the whole infinite chain. This is apparently
quite a non-local effect, but it can in fact be converted to a local
interaction provided the end-to-end vector is known.

Suppose we have end-to-end vector ${\bf v}$, then if we know that ${\bf
x}$ is a site on the walk then necessarily all sites $y = {\bf x} + j
{\bf v}, j \in \Z$ must also be part of the endless walk. We should
therefore test for self-intersections for an endless walk $\omega$ by
checking for self-intersections in reduced coordinates 
$\omega(i) / {\bf v}$.

Our algorithm generated all eSAW with a given end-to-end vector, ${\bf
v}$, and used reduced coordinates to test for self-intersections.
Starting at the origin, we incrementally grow our walk by considering
all possible ways of adding an additional step which are compatible with
the end-to-end vector. As the walk is extended we add $\omega(i)$ to
the list of visited sites, and we check to see if $\omega(i) / {\bf v} =
\omega(j) / {\bf v} $ for any $j < i$. If there is an intersection we
know that this sub-walk can never be a valid eSAW, and so we truncate
the backtracking tree.

The reduced coordinates of visited sites were stored in a
hash table to ensure that intersection testing took $O(1)$ CPU time.
Optimal backtracking
implementations take CPU time which is of exactly the same order as the
number of objects being generated, and in this sense our implementation
was optimal. It certainly would be possible to obtain a small constant
factor improvement in performance through fine tuning the implementation.

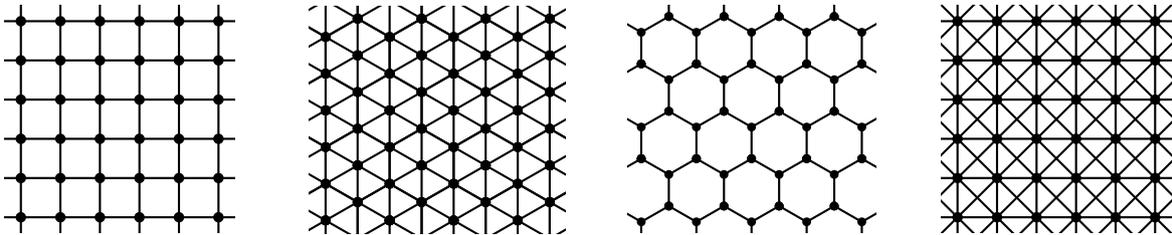
\begin{figure}[htb]
\begin{center}
\begin{tikzpicture}[thick,scale=0.52]
\begin{scope}[shift={(0,-1.4)}]
\clip (-0.4,1.4) rectangle (5.4,-4.4);
\foreach \i in {0,...,5} {
\draw (\i,2) -- (\i,-5);
}
\foreach \j in {1,...,-4} {
\draw (-1,\j) -- (6,\j);
}
\foreach \i in {0,...,5} {
\foreach \j in {1,...,-4} {
\draw[fill] (\i,\j) circle (3pt);
}
}
\end{scope}
\begin{scope}[shift={(6.9,0.60)},scale=0.93548]
\clip (0.4330127,-0.65) rectangle (7.361216,-6.85);
\foreach \j in {0,-3.,-6.} {
\foreach \i in {0,1.732051,3.464102,5.1961524,6.9282,8.660254} {
\draw[fill] (\i,\j) circle (3pt);
\draw[fill] (\i,\j) \d circle (3pt) \ew
\draw[fill] (\i,\j) \lu circle (3pt) \ew
\draw[fill] (\i,\j) \ru circle (3pt) \ew
\draw[fill] (\i,\j) \u circle (3pt) \ew
\draw[fill] (\i,\j) \ld circle (3pt) \ew
\draw[fill] (\i,\j) \rd circle (3pt) \ew
}
}
\foreach \j in {-1.,-4.,-7.} {
\foreach \i in {0,1.732051,3.464102,5.1961524,6.9282,8.660254} {
\draw[fill] (\i,\j) circle (3pt);
\draw[fill] (\i,\j) \d circle (3pt) \ew
\draw[fill] (\i,\j) \lu circle (3pt) \ew
\draw[fill] (\i,\j) \ru circle (3pt) \ew
\draw[fill] (\i,\j) \u circle (3pt) \ew
\draw[fill] (\i,\j) \ld circle (3pt) \ew
\draw[fill] (\i,\j) \rd circle (3pt) \ew
}
}
\foreach \j in {-1.5,-4.5,-7.5} {
\foreach \i in {0.8660254,2.598076,4.330127,6.0621778265,7.794228634} {
\draw[fill] (\i,\j) circle (3pt);
\draw[fill] (\i,\j) \d circle (3pt) \ew
\draw[fill] (\i,\j) \lu circle (3pt) \ew
\draw[fill] (\i,\j) \ru circle (3pt) \ew
\draw[fill] (\i,\j) \u circle (3pt) \ew
\draw[fill] (\i,\j) \ld circle (3pt) \ew
\draw[fill] (\i,\j) \rd circle (3pt) \ew
}
}
\foreach \j in {-2.5,-5.5,-8.5} {
\foreach \i in {0.8660254,2.598076,4.330127,6.0621778265,7.794228634} {
\draw[fill] (\i,\j) circle (3pt);
\draw[fill] (\i,\j) \d circle (3pt) \ew
\draw[fill] (\i,\j) \lu circle (3pt) \ew
\draw[fill] (\i,\j) \ru circle (3pt) \ew
\draw[fill] (\i,\j) \u circle (3pt) \ew
\draw[fill] (\i,\j) \ld circle (3pt) \ew
\draw[fill] (\i,\j) \rd circle (3pt) \ew
}
}
\end{scope}
\begin{scope}[shift={(15,0.525)},scale=0.80555555]

\clip (0.4330127,-0.65) rectangle (8.227241336,-7.85);
\foreach \j in {0,-3.,-6.} {
\foreach \i in {0,1.732051,3.464102,5.1961524,6.9282,8.660254} {
\draw[fill] (\i,\j) circle (3pt);
\draw[fill] (\i,\j) \d circle (3pt) \ew
\draw[fill] (\i,\j) \lu circle (3pt) \ew
\draw[fill] (\i,\j) \ru circle (3pt) \ew
}
}
\foreach \j in {-1.5,-4.5,-7.5} {
\foreach \i in {0.8660254,2.598076,4.330127,6.0621778265,7.794228634} {
\draw[fill] (\i,\j) circle (3pt);
\draw[fill] (\i,\j) \d circle (3pt) \ew
\draw[fill] (\i,\j) \lu circle (3pt) \ew
\draw[fill] (\i,\j) \ru circle (3pt) \ew
}
}
\end{scope}
\begin{scope}[shift={(23.7,-1.4)},inner sep=0pt,outer sep=0pt]
\clip (-0.4,1.4) rectangle (5.4,-4.4);
\foreach \i in {0,...,5} {
\draw (\i,2) -- (\i,-5);
}
\foreach \j in {1,...,-4} {
\draw (-1,\j) -- (6,\j);
}
\foreach \i in {-10,...,5} {
\draw (\i,2) -- (\i+8,-6);
\draw (\i,-6) -- (\i+8,2);
}
\foreach \i in {0,...,5} {
\foreach \j in {1,...,-4} {
\draw[fill] (\i,\j) circle (3pt);
}
}
\end{scope}
\end{tikzpicture}
\end{center}
\caption{Two-dimensional lattices, from left to right: square,
triangular, honeycomb, and union jack.\label{fig:lattices}}
\end{figure}

We performed our enumerations on the following lattices in two
dimensions: square, triangular, honeycomb, and union jack, and in three
dimensions: simple cubic, body centered cubic (bcc) and face centered
cubic (fcc). 
The two-dimensional lattices are shown in Fig.~\ref{fig:lattices}.
The lattice we refer to as ``union jack'' is non-standard,
as it allows for crossings, and may be properly described as a square
lattice with additional bonds connecting next-to-nearest
neighbors~\cite{Oitmaa1980,Narayanan1995}.

We enumerated $e_n$ and $er_n$ directly, and the fact that the resulting
value for $d_n$ must be
an integer places quite a strong consistency condition on our
enumerations.
Another condition that our enumerations should satisfy comes from the
observation that eSAW form equivalence classes under
transformations of the underlying lattice. For the square lattice, eSAW which
explore both available dimensions must form an equivalence class under
the full symmetry group, $O(2)$, which means that $e_n - 4$
must be divisible by 8. The corresponding condition for the simple
cubic lattice is that
$e_n^{\text{(simple cubic)}} - 3e_n^{\text{(square)}} +6$ must be divisible by 48, the number of
elements of $O(3)$. 
We have confirmed that each of these
conditions hold for our data.

Interestingly, SAC do not necessarily exhibit the full symmetry group of
the underlying lattice. The symmetries under which SAC are invariant is
determined purely by the end-to-end vector.

The combination of integer values for $d_n$, correct symmetry factors
for $e_n$, and correct low order values gives us confidence that the
enumerations reported in 
Tables~\ref{tab:z2}--\ref{tab:honeycomb} in \ref{sec:data}
are correct.
As an aside, we note that for $n$ prime, the sequences must satisfy 
\begin{alignat}{4}
e_{2n} &= i_{2n} + i_2 = 2 n d_{2n} + 2 d_2 &\hspace{2cm}
&\text{honeycomb lattice,}
\\ e_n &= i_n + i_1 = 2 n d_n + 2 d_1 &\hspace{2cm} &\text{all other
lattices.}
\end{alignat}

We performed our enumerations on a SunFire X4600M2 with 2.3GHz AMD
Opteron CPUs. 
For each lattice we divided the backtracking enumeration amongst 20
different processors. The total number of CPU hours required for the enumerations
was: 12000 for the square lattice, 2100 for the
triangular lattice, 2400 for the honeycomb lattice, 2000 for the union jack
lattice, 5600 for the simple cubic lattice, 7000 for the bcc lattice,
and 1300 for the fcc lattice.

Our enumeration data are collected in
Tables~\ref{tab:z2}--\ref{tab:honeycomb} in \ref{sec:data}.
Note that we only report $r_n$ for the square and simple cubic lattices.
A bug in our enumeration program for the other lattices 
meant that the data we collected suffered from rounding
errors. These data would allow us to
estimate the amplitude $\De$, but since it is a non-universal quantity we
regard $\De$ to be of relatively minor interest. We will present
enumerations of $r_n$ for other lattices in future papers.

Finally, note that we do not regard basic backtracking as best practice
for performing the enumerations, rather as a sensible first step towards
studying this new model. We briefly mention alternative enumeration
approaches in the following subsection.

\subsection{Alternative enumeration methods\label{sec:alternativeenumeration}}

More sophisticated enumeration algorithms hold promise for future
progress, but implementing them for eSAW will require overcoming
certain challenges
due to the apparent non-locality of
eSAW self-interactions. Our intuition is that this may present a
significant technical challenge for the approaches mentioned below, but
not an insurmountable obstacle.

The most straightforward approach to improving our enumerations would be
to implement the two-step method~\cite{Clisby2007}. This could certainly
be used to extend the series in \ref{sec:data} by a few terms
in each case for the same computational effort. We are not sure if the
lace expansion technique described in the same paper could be useful 
for enumerations of eSAW.

The most powerful known method for three-dimensional walks is the
length-doubling algorithm~\cite{Schram2011,Schram2013}. It is not clear to us how this
exciting new enumeration algorithm could be applied to eSAW.

We believe the most promising approach is the finite lattice
method~\cite{deNeef1977,Enting1980}, which has been applied to the
problem of enumerating SAW~\cite{Jensen2004} and SAP~\cite{Clisby2012}
on two-dimensional lattices. A recent improvement to the finite
lattice method~\cite{Clisby2012} raises the possibility that the
method could also be usefully applied in three dimensions.

\section{Analysis\label{sec:analysis}}

Our goal in analyzing the series from Appendix \ref{sec:data} is to
extract as much information as possible. In particular we wish to
provide numerical support for our argument that $\gammae = 1$ in
Sec.~\ref{sec:gamma_analysis} and $\nue = \nu$ in
Sec.~\ref{sec:nu_analysis}. In Sec.~\ref{sec:amplitude_analysis} we
provide evidence for the surprising fact that
the amplitude $\Ae$ is universal, while confirming that there are no
analytic corrections to scaling for the number growth.

For our analyses we use arguably the two most powerful, general purpose
methods of series analysis available: the methods of direct fitting and
differential approximants. We succinctly describe each of these methods
below in Sections~\ref{sec:direct} and \ref{sec:DA} respectively.

In our analyses we will ruthlessly exploit our prior knowledge of
critical parameters, in particular the values of the connective
constants of the various lattices.  For two-dimensional lattices, the
best estimates for $\mu$ come from series derived from the finite
lattice method. For the square lattice, $\mu =
2.638\,158\,530\,35(2)$~\cite{Clisby2012}, while for the triangular
lattice $\mu = 4.150\,797\,226(26)$~\cite{Jensen2004a}. For the
honeycomb lattice we have the exact result $\mu = \sqrt{2 + \sqrt{2}} =
1.847\,759\,065\,022\cdots$~\cite{Nienhuis1982,Duminil-Copin2012b}.  For
the union jack lattice there are no accurate prior estimates of the
connective constant, and so in contrast to the other series we perform
unbiased fits.  From a recent Monte Carlo calculation, we know that $\mu
= 4.684\,039\,931(27)$~\cite{Clisby2013} for the simple cubic lattice.
Using the length-doubling algorithm, Schram~\cite{Schram2011b} has
derived series for the bcc and fcc lattices to 28 and 24 steps
respectively. A preliminary analysis of these data leads to $\mu =
6.530\,521(6)$ (bcc) and $\mu = 10.037\,067(7)$ (fcc).

The values above for the connective constants are in all cases
sufficiently accurate that we can consider them to be ``exact'', i.e. we
do not have to measure the effect of the error in the estimates of $\mu$
on our fits. This is not usually the case with biased fits. Our
situation is special because this knowledge comes from other sources,
and $\mu$ is known to much higher precision than could possibly be
obtained from our short eSAW series. We have tested this assumption for
ourselves, but will not discuss this issue any further in the analysis
below.

We found that eSAW series are in every case somewhat better behaved than
the irreducible eSAW series. Generically, they are quite different to
begin with but then the fits converge for larger $n$, since they are
asymptotically identical.  We attribute this to the fact that for small
$n$ the effect of repeating sub-walks for composite $n$ values is quite
high, whereas when $n$ is large this effect becomes negligible.  To
reduce visual noise only the eSAW series are plotted.

We are mindful of the following observation due to
Guttmann~\cite{Guttmann1989} with reference to extrapolations from
series analysis: ``error bounds are generally referred to
as (subjective) confidence limits, and as such frequently measure the
enthusiasm of the author rather than the quality of the data.''. Where
we give an explicit confidence interval for our estimates we will try to
moderate our enthusiasm. We will also present all of our fits in
graphical form so
that readers can make their own judgments.

\subsection{Direct fitting\label{sec:direct}}

The method of direct fitting involves fitting the asymptotic form for the series with
the series coefficients. This approach is especially useful for finding
amplitudes. 

The idea is to truncate
(\ref{eq:enasympt})--(\ref{eq:erhonasympt}) at some finite
order, and then fit a subset of coefficients from the corresponding
series to this truncation. 
Rather than going in to any great depth about the general method, we
will instead explain exactly what fits are performed in the analysis
below. The most detailed explanation is in
Sec.~\ref{sec:gamma_analysis}.

See \cite{Guttmann1989,Clisby2007,Guttmann2009} for detailed
explanations of the direct fitting approach.

\subsection{Differential approximants\label{sec:DA}}
In the method of differential approximants,
the unknown generating function is represented by the solution to an ordinary
differential equation of the form:
\begin{align}
\sum_{i=0}^{K} Q_i(z) \left(z \frac{d}{dz}\right)^i f(z) &= P(z).
\end{align}
The functions $P$ and $Q_i$ are polynomials, of degrees $L$ and $N_i$.
We choose $0 \leq L \leq 10$, $N_K \geq 3$ (which guarantees at least
three regular singularities), and take
$Q_K$ to have highest-order coefficient equal to $1$.
We choose $K = 2$ for the two-dimensional lattices, and $K = 1$ for the
three-dimensional lattices as these series are generally shorter.
The order of the polynomials was chosen so that $|N_i - N_j| \leq 3$.
Given coefficients $a_0,\ldots, a_N$, the polynomials $P,Q_i$ are chosen
so that the polynomial $\sum_{n=0}^Na_nz^n$ solves the differential equation to
within an error of order $z^{N+1}$. This choice is made by solving
a system of linear equations in $L+K+1+\sum_{i=0}^K N_i$ unknowns,
determined from $N+1$ known coefficients.

Once the approximant for a given choice of $L$, $K$, and $N_i$ has been
determined, one may then find the singularities of the differential
equation by determining the zeroes of $Q_K(z)$. The exponent
corresponding to each regular singularity is then determined from the
indicial equation. For the singularity $z_i$, the corresponding exponent
is:
\begin{align}
\lambda_i &= K - 1 - \frac{Q_{K-1}(z_i)}{z_i Q^\prime(z_i)}.
\end{align}

The physically relevant singularity is the nearest singularity on the
positive real axis. If we denote this singularity as $z_c$, then the
connective constant $\mu = 1/z_c$. We denote approximants which have
singularities near the physical singularity, or clearly incorrect
singularities on or near the positive real axis, as ``defective''. To
estimate $z_c$ and the corresponding critical exponent we determine the
leading singularity for a large number of choices of $L$ and $N_i$, and
prune defective approximants away.
We then plot these approximant estimates of the critical exponent
against $n$ (the power of the highest order series coefficient used) in order to extrapolate and obtain a final best estimate for
the critical exponents $\gammae$ and $\nue$.

If desired, 
it is straightforward to ensure that there is a
singularity at a biased value of $z_c$ by introducing an additional
linear equation:
\begin{align}
Q_K(z_c) &= 0.\label{QK0}
\end{align}
In this paper, we are able utilize accurate estimates for $\mu$ from
longer SAW series, and so all of our differential approximant analyses
are biased.

See Guttmann~\cite{Guttmann1989} for far more information about
differential approximants and series analysis, or Guttmann and
Jensen~\cite{Guttmann2009} for a recent overview.

\subsection{$\gammae$\label{sec:gamma_analysis}}

\begin{figure}[!htb]
\begin{center}
\includegraphics[scale=0.8]{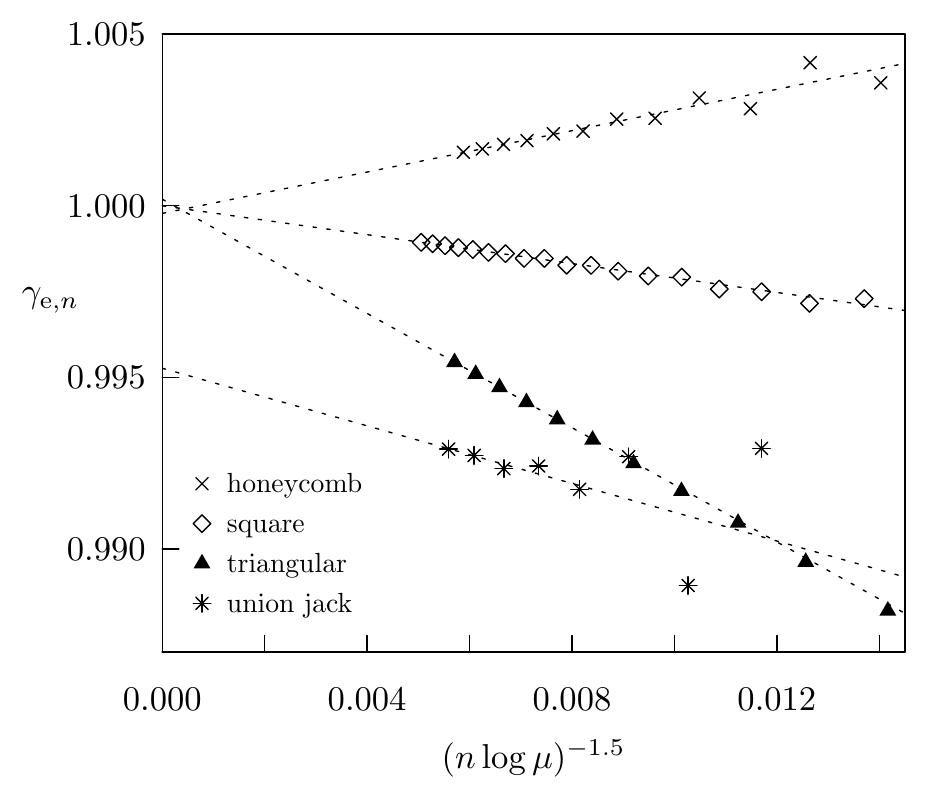}
\includegraphics[scale=0.8]{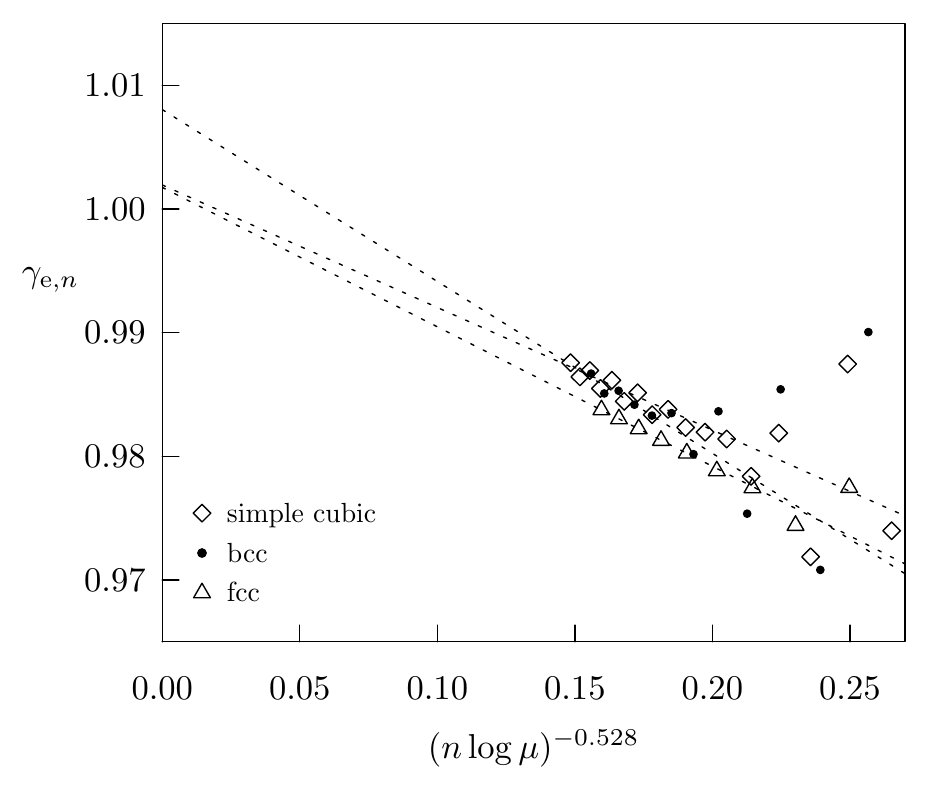}
\vspace{-3ex}
\end{center}
\caption{Direct fit estimates of $\gammae$ for two-
 and three-dimensional lattices (left and right respectively).\label{fig:gamma_direct}}
\end{figure}

To determine $\gammae$ for the square, honeycomb, and triangular
lattices we take the logarithm of $e_n/\mu^n$, and note that the
asymptotic form we derive from (\ref{eq:enasympt}) is
\begin{align}
\log \left(e_n/\mu^n\right) &\sim ({\gammae - 1}) \log n + \log \Ae + \left[\frac{a_1}{n} +
\frac{b_0}{n^{\Delta_1}} \right].
\end{align}
We single out the leading order neglected terms by placing them in large
square brackets.  Note that we have substituted $\mu$ for $\mue$ due to
the proof in Sec.~\ref{sec:proof}.  We then solve for $\gammae$ and
$\log \Ae$ by using terms from the sequence. For the triangular and
honeycomb lattices we use consecutive terms, say $e_{k-1}$ and $e_k$ for
the triangular lattice and $e_{k-2}$ and $e_k$ for the honeycomb
lattice. For the square lattice we find that we obtain smoother
fits by fitting alternate terms $e_{k-2}$ and $e_k$. We label the
resulting estimates for $\gammae$ as $\gammaek$, and plot these
estimates in Fig.~\ref{fig:gamma_direct}.  Note that we choose to take
the logarithm in the above equation simply as a matter of convenience,
so that the equation to be solved for $\gammae$ and $\log \Ae$ is
linear.

Now, we expect that the error term in our $\gammaen$ fits is of the
order of the first neglected term. 
The asymptotic behavior of two-dimensional SAW is 
well understood~\cite{Caracciolo2005}, thanks to a combination of series
analysis of long series generated via the finite lattice method, and high
powered Monte Carlo simulations.
This leads us to believe that
$\Delta_1 = 3/2$ exactly. 
The argument of Sec.~\ref{sec:analytic} suggests that $a_1 = 0$, and so we
expect that 
plots of
$\gammaen$ versus $n^{-\Delta_1}$ should be linear, and this is
confirmed for each lattice in Fig.~\ref{fig:gamma_direct}.
We confirmed that 
plots of
$\gammaen$ versus $1/n$ possess significant curvature, indicating that
$a_1$ must be close to zero in each case; we will estimate $a_1$ in
Sec.~\ref{sec:amplitude_analysis}.

We employ one additional trick in our plots of direct fit estimates. For
a lattice with connective constant $\mu$, when we plot against a
remainder of the form $n^{-a}$ we normalize it as $x = (n \log \mu)^{-a}$.
This ensures that 
for any given value of $x$ the number of eSAW is
approximately the same, independent of the lattice. Loosely speaking, we
want the $x$ coordinate to be correlated with the information content of
the series.

We also obtain direct fits for $\gammae$ from the union jack
lattice, solving
\begin{align}
\log e_n &\sim n \log \mu + ({\gammae - 1}) \log n + \log \Ae + \left[\frac{a_1}{n} +
\frac{b_0}{n^{\Delta_1}} \right]
\end{align}
for $\log \mu$, $\gammae$, and $\log \Ae$ using coefficients $e_{k-2}$,
$e_{k-1}$, and $e_k$. These estimates are not as
well behaved because we do not have the benefit of a prior high
precision estimate for $\mu$.

Examining Fig.~\ref{fig:gamma_direct} it is clear that all of the
estimates for $\gammaen$ are converging smoothly to $1$, as predicted.
As a visual guide we fit a straight line to the five highest order
estimates for each lattice, and their y-intercepts are extremely close
to 1, especially for our high precision data from the square, honeycomb
and triangular lattices.
From the figure, $\gammae$ must be within the range $1.000(1)$. We could
do better with the direct fits, but we will see that the differential
approximant analysis below is more than adequate for the task of placing
a tight bound on $\gammae$.

For the three-dimensional lattices, we once again fit the asymptotic
form
\begin{align}
\log \left(e_n/\mu^n\right) &\sim ({\gammae - 1}) \log n + \log \Ae + \left[\frac{a_1}{n} +
\frac{b_0}{n^{\Delta_1}} \right].
\end{align}
For three-dimensional SAW, the asymptotic form is well understood
theoretically, but the numerics are far more difficult.
In~\cite{Clisby2007} there is extensive discussion of the asymptotic
form, while in \cite{Clisby2010} there is the most detailed and accurate
numerical investigation of corrections to scaling that is currently
available. For three-dimensional lattices we have
$\Delta_1=0.528(12)$~\cite{Clisby2010} as the dominant neglected term,
but in other respects we proceed in the same manner as for the
two-dimensional lattices. For the simple cubic and bcc lattices we fit
$e_{k-2}$ and $e_k$, while for the fcc lattice we fit $e_{k-1}$ and
$e_k$.

Our fits for $\gammae$ are shown in Fig.~\ref{fig:gamma_direct}. It is
clear in each case that the estimates are smoothly converging to
$\gammae = 1$. The fits as shown suggest that $\gammae = 1.005(10)$,
which encompasses $\gammae = 1$ as predicted.
We do need to consider the effect of the known error in $\Delta_1 =
0.528(12)$, and so re-performed the analysis for $\Delta_1 = 0.516$ and
$\Delta_1 = 0.54$. In our case, because $\Delta_1$ is known to high
precision from another source, this does not materially change our
estimate for $\gammae$.

As a technical aside, we find that for a number of lattices,
particularly the square lattice, the simple cubic lattice, the bcc
lattice, and to a lesser extent the honeycomb lattice, there is quite a
high degree of oscillation with our direct fit estimates.  We reduced
the amplitude of the oscillations by fitting alternate terms, and on some
occasions we also smoothed the fits by averaging consecutive estimates.

As mentioned in the description of the asymptotic forms
(\ref{eq:cnasympt})--(\ref{eq:rhonasympt}), for SAW there is a
contribution to the asymptotic form from the so-called
anti-ferromagnetic singularity. In the analyses performed in this paper
we avoid having to deal with this singularity directly by treating the
series arising from odd and even terms separately. As explained 
in Sec.~\ref{sec:af}, there is no clear evidence as to whether the
oscillations in the series are strictly odd-even, which would be a
signal of the anti-ferromagnetic singularity on the negative real axis.
The approximate alternation of some series is \emph{not} conclusive, as
it may be due to singularities near to but not on the negative real
axis. Instead, our choice to average consecutive terms and fit
alternating terms is based purely on the observation that this makes the
fits easier to extrapolate.

\begin{figure}[!htb]
\begin{center}
\includegraphics[scale=0.8]{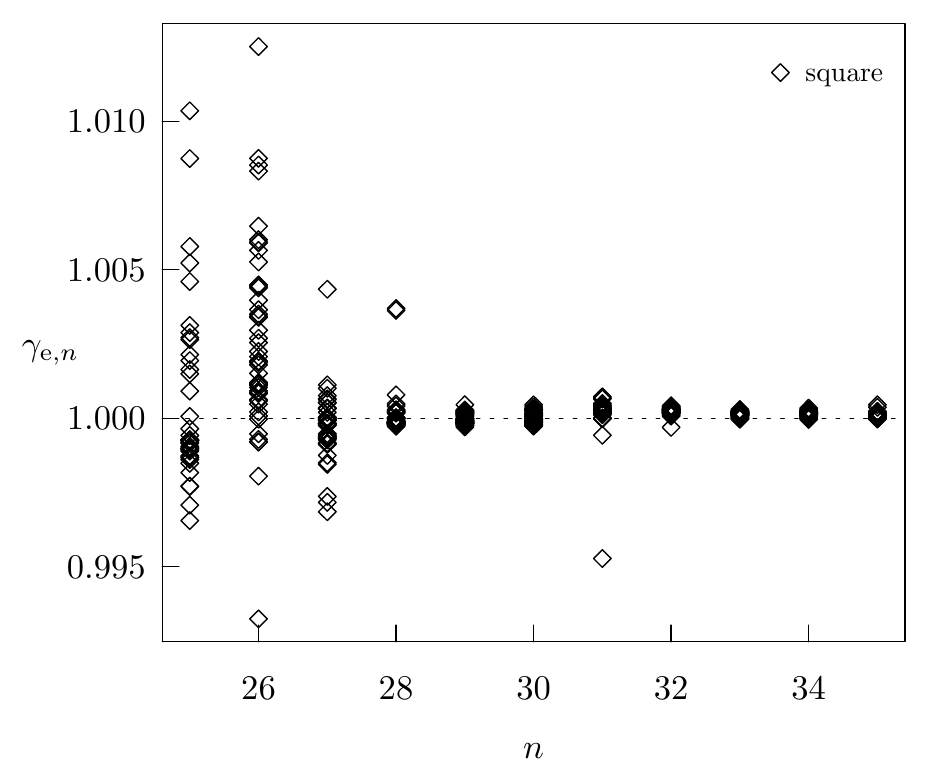}
\includegraphics[scale=0.8]{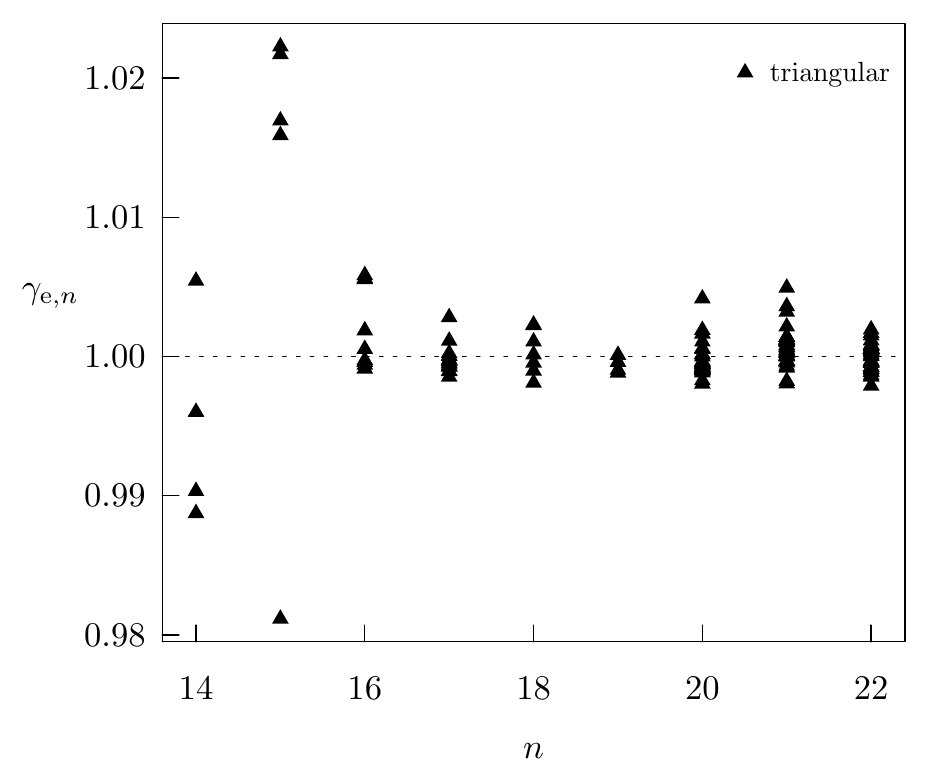}
\vspace{-3ex}
\end{center}
\caption{Differential approximant estimates of $\gammae$ on the
square (left) and triangular (right) lattices.\label{fig:gamma_DAe2}}
\end{figure}

\begin{figure}[!htb]
\begin{center}
\includegraphics[scale=0.8]{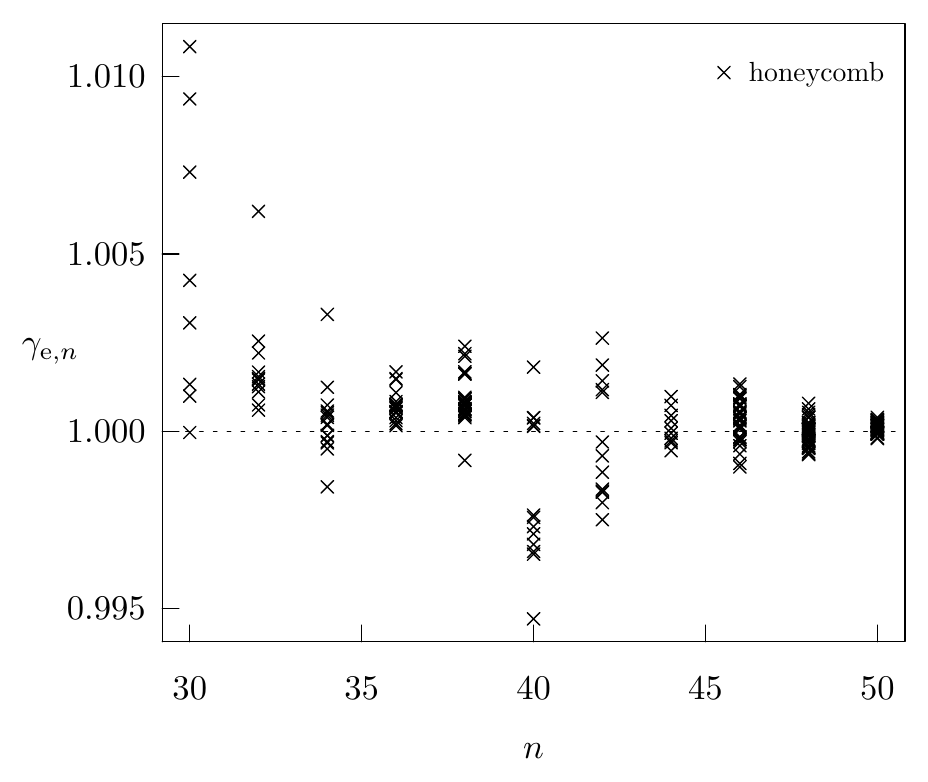}
\includegraphics[scale=0.8]{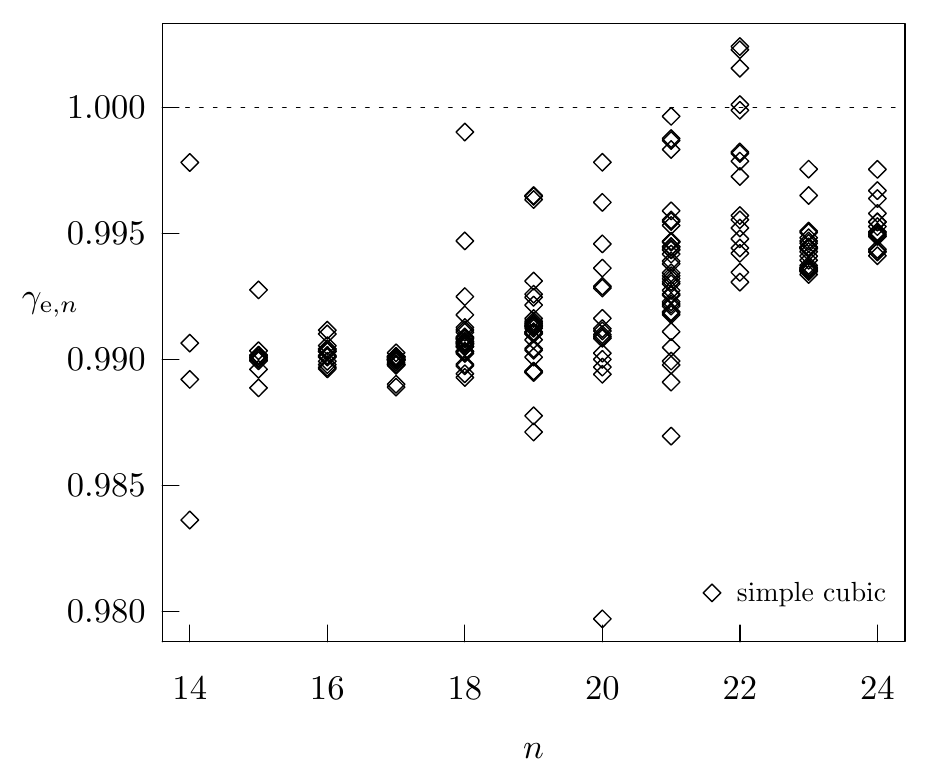}
\vspace{-3ex}
\end{center}
\caption{Differential approximant estimates of $\gammae$ on the
honeycomb (left) and simple cubic (right)
lattices.\label{fig:gamma_DAehc}}
\end{figure}

\begin{figure}[!htb]
\begin{center}
\includegraphics[scale=0.8]{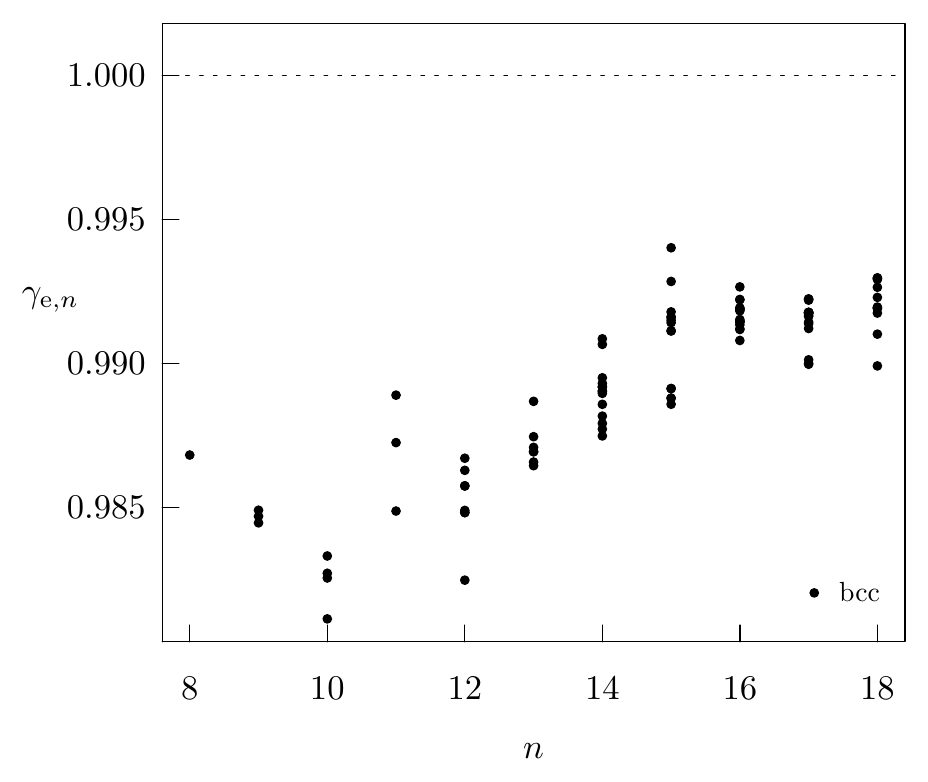}
\includegraphics[scale=0.8]{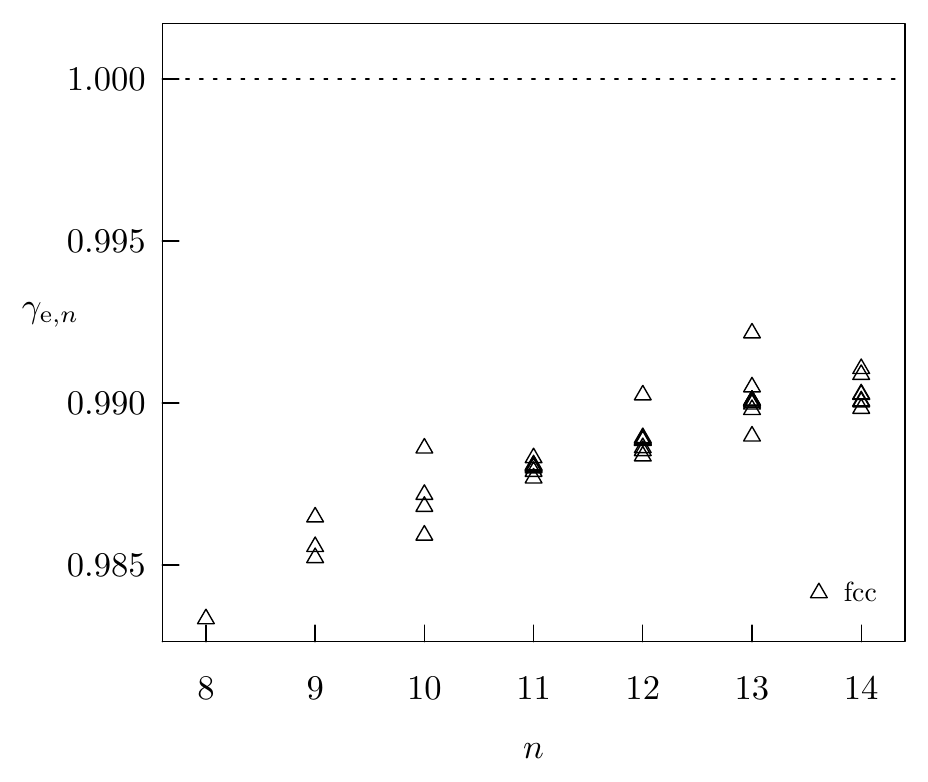}
\vspace{-3ex}
\end{center}
\caption{Differential approximant estimates of $\gammae$ on the
bcc (left) and fcc (right) lattices.\label{fig:gamma_DAebcc}}
\end{figure}

We now proceed to estimate $\gammae$ for the various lattices via the
differential approximant method described in
Sec.~\ref{sec:DA}. Our fits are reproduced in
Figures~\ref{fig:gamma_DAe2}, \ref{fig:gamma_DAehc}, and
\ref{fig:gamma_DAebcc}.


For the two-dimensional lattices, the differential approximants appeared
to converge better for the square and honeycomb lattices. Nonetheless,
combining information from all three suggests that $\gammae = 1.000(1)$
in line with our direct fit estimate.

For the three-dimensional lattices the estimates are not as well
converged, but this is to be expected because the series are shorter,
and the leading correction to scaling is larger. Nonetheless, the trend
in each case seems to be converging to one, and suggests that $\gammae =
0.99(2)$. We acknowledge that extrapolating such estimates is a fraught
process, especially since there is clearly still systematic drift in the
estimates. In addition, we already `know' the answer, and this can
easily introduce a subconscious bias. However, our interest here is less
in deriving an estimate for $\gammae$, than in showing that $\gammae =
1$ is the most plausible interpretation of the available evidence.

We consider the series analyses presented to be quite
convincing in their support of our prediction that $\gammae = 1$ for
eSAW on two- and three-dimensional lattices.
For the reminder of the analysis section we will \emph{assume} that
$\gammae = 1$, and use this information to bias our fits.

\FloatBarrier

\subsection{$\nue$\label{sec:nu_analysis}}
For the purpose of estimating $\nue$ we find that the $er_n$ series is
far more useful than the $e\rho_n$ series.
We only have reliable data for the square and simple cubic lattices, but
we given that $\nue$ is a universal quantity this is not a problem.

\begin{figure}[!htb]
\begin{center}
\includegraphics[scale=0.8]{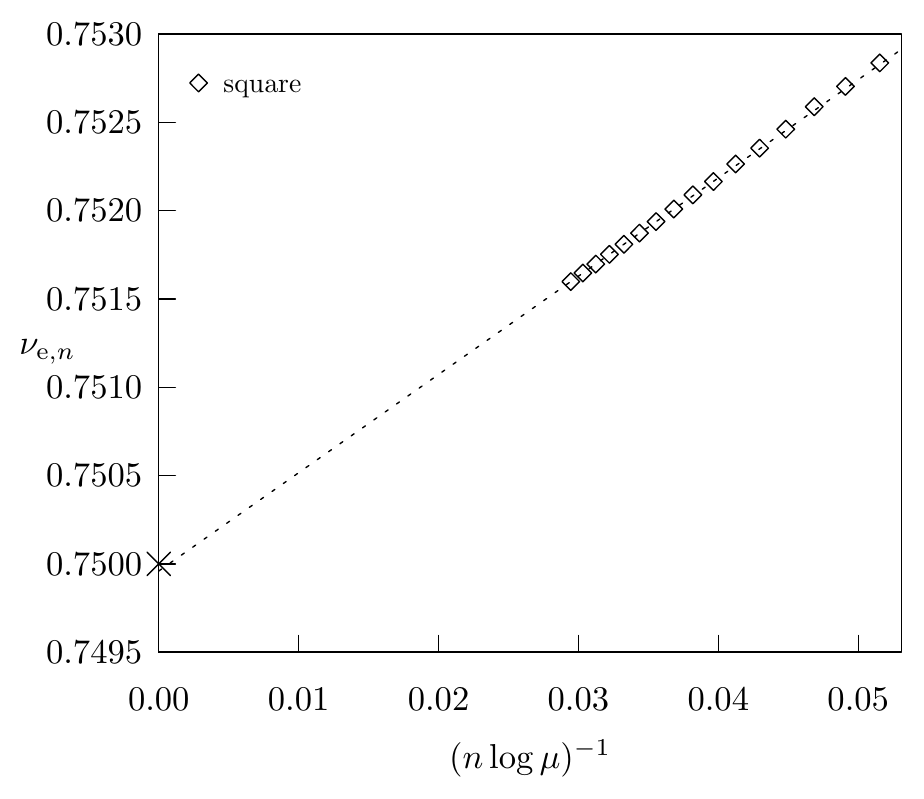}
\includegraphics[scale=0.8]{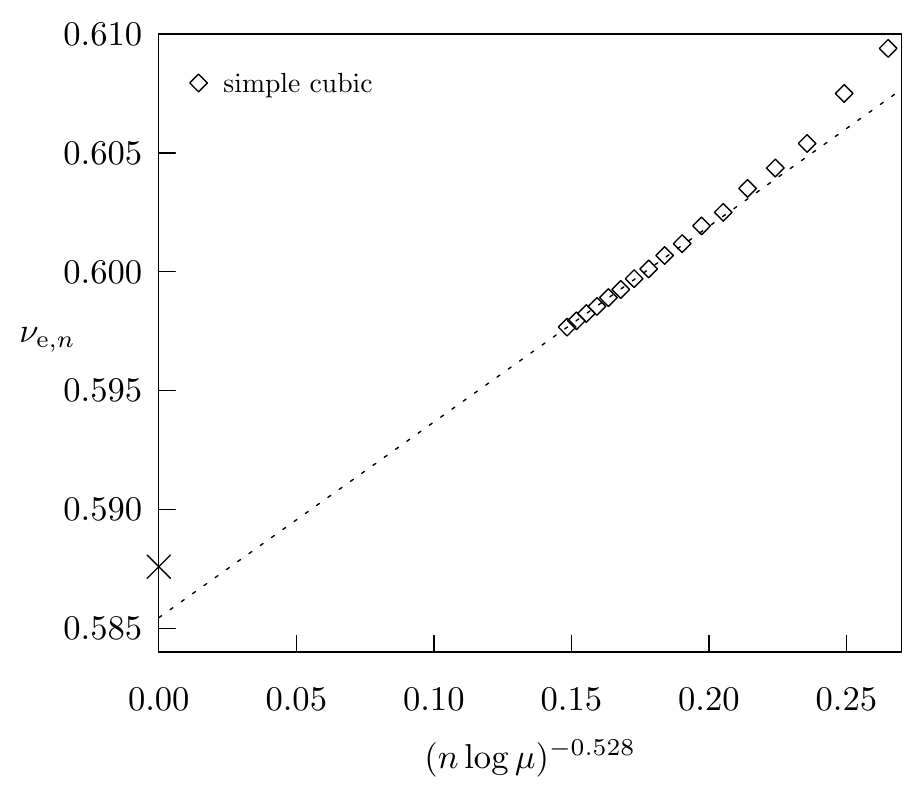}
\vspace{-3ex}
\end{center}
\caption{Direct fit estimates of $\nue$ for 
the square (left) and simple cubic (right) lattices.\label{fig:nu_direct}}
\end{figure}

We assume that $\gammae = 1$, and substitute this into the asymptotic
expression for $er_n$, and then perform direct fits on the following
asymptotic form:
\begin{align}
\log (er_n/\mu^n) &\sim  ({2\nue + \gammae - 1}) \log n +  \log(\Ae \De) + \left[
\frac{f_1}{n} + \frac{g_0}{n^{\Delta_1}} \right],
\\ &\sim  {2\nue} \log n +  \log(\Ae \De) + \left[
\frac{f_1}{n} + \frac{g_0}{n^{\Delta_1}} \right].
\end{align}
For each of the lattices we fit alternating terms, $e_{k-2}$ and $e_k$,
to obtain estimates for $\nuek$ and the amplitude, and we plot the
results in Fig.~\ref{fig:nu_direct}.

For the square lattice, it is very clear that the leading analytic
correction to scaling has non-zero amplitude, as shown by the linear
behavior of the fits in Fig.~\ref{fig:nu_direct}. The convergence to $\nue =
\nu = 3/4$ is extremely convincing; we conservatively interpret our
results as suggesting $\nue = 0.7500(1)$.

For the simple cubic lattice, the convergence is not as good, but this
is to be expected for such a short series with strong corrections to
scaling. Our estimates in Fig.~\ref{fig:nu_direct} suggest that $\nue$
is in the vicinity of $0.586$, while the best estimate for $\nu$, $\nu =
0.587\,597(7)$~\cite{Clisby2010}, is shown on the plot.

\begin{figure}[!htb]
\begin{center}
\includegraphics[scale=0.8]{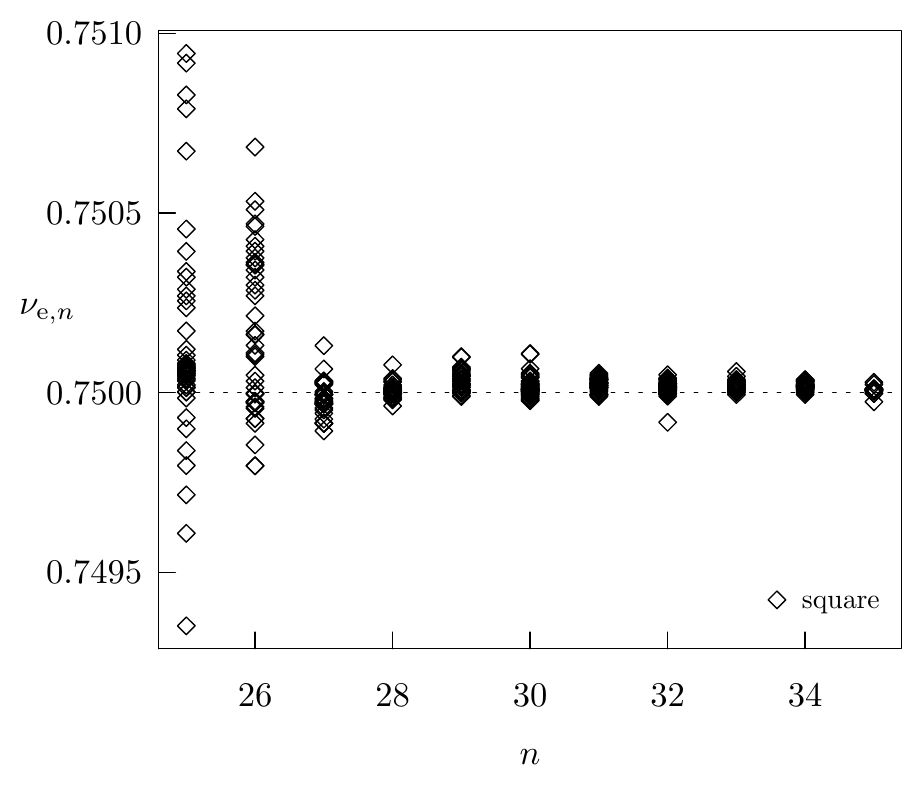}
\includegraphics[scale=0.8]{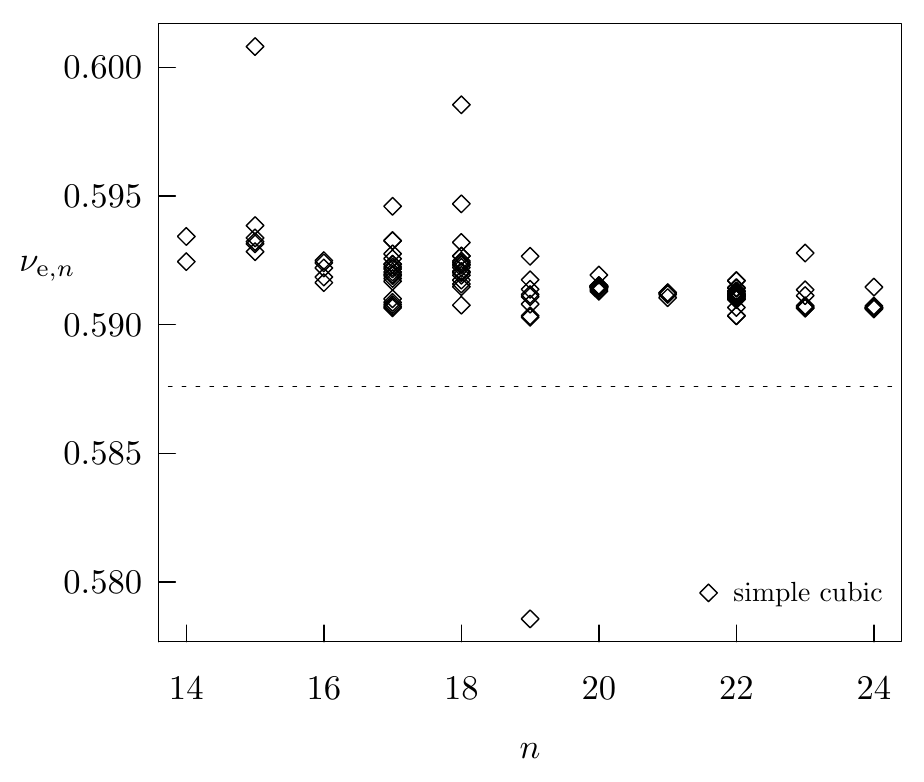}
\vspace{-3ex}
\end{center}
\caption{Differential approximant estimates of $\nue$ on the
square (left) and simple cubic (right) lattices.\label{fig:nu_DA}}
\end{figure}

The results from our differential approximant analyses are shown in
Fig.~\ref{fig:nu_DA}. Technically, we are in fact plotting estimates of
$\nue + (\gammae - 1)/2$, but we assume that $\gammae = 1$ as stated
previously.

For the square lattice the convergence of the differential approximants
is very good, and $\nue = 0.7500(1)$ is a conservative interpretation of
the data.

For the simple cubic lattice, once again there is still a
systematic drift occurring at $n=24$, and the estimates for $\nue$ are
approximately clustered around 0.590-0.593. This is consistent with
$\nue = \nu$, but not completely convincing.

We believe that the numerical evidence supporting $\nue = \nu$ is
unequivocal for the square lattice. For the simple cubic lattice we
cannot be so definite due to the short series available, but we
nonetheless strongly believe that $\nue = \nu$ is the most plausible
interpretation of the evidence for three-dimensional lattices also.

To obtain definitive evidence in support of $\nue = \nu$ in three
dimensions we will either need to significantly extend the series, or
apply the pivot algorithm~\cite{Clisby2010}. The pivot algorithm in the
context of eSAW is discussed in Sec.~\ref{sec:discussion}.

\FloatBarrier

\subsection{$\Ae$\label{sec:amplitude_analysis}}

\begin{figure}[!htb]
\begin{center}
\includegraphics[scale=0.8]{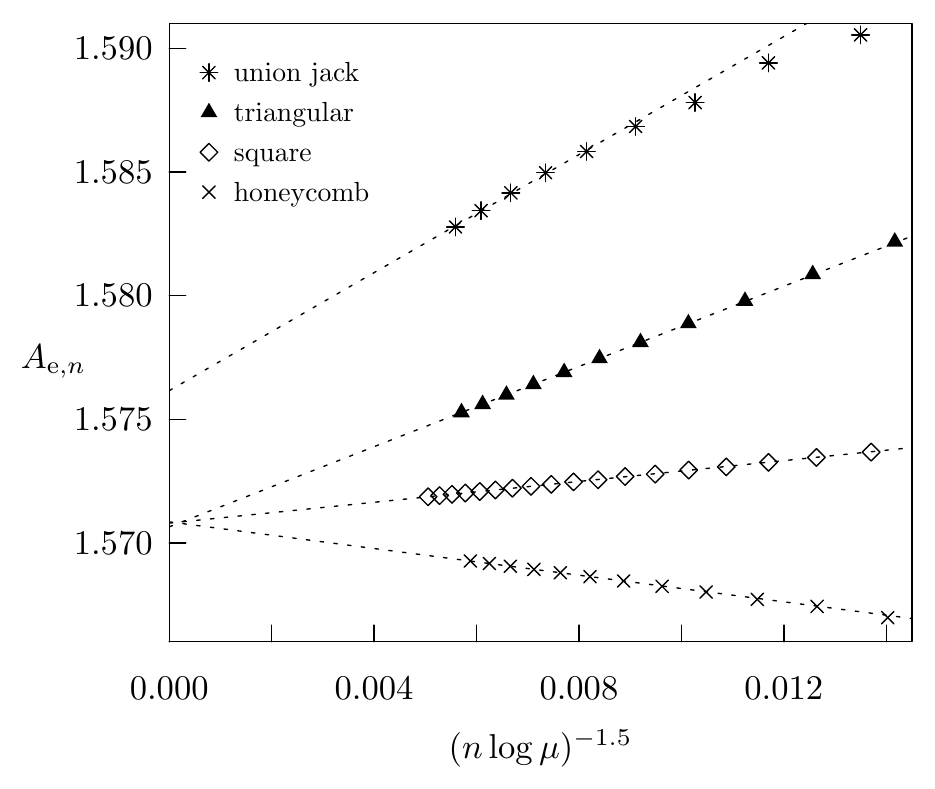}
\includegraphics[scale=0.8]{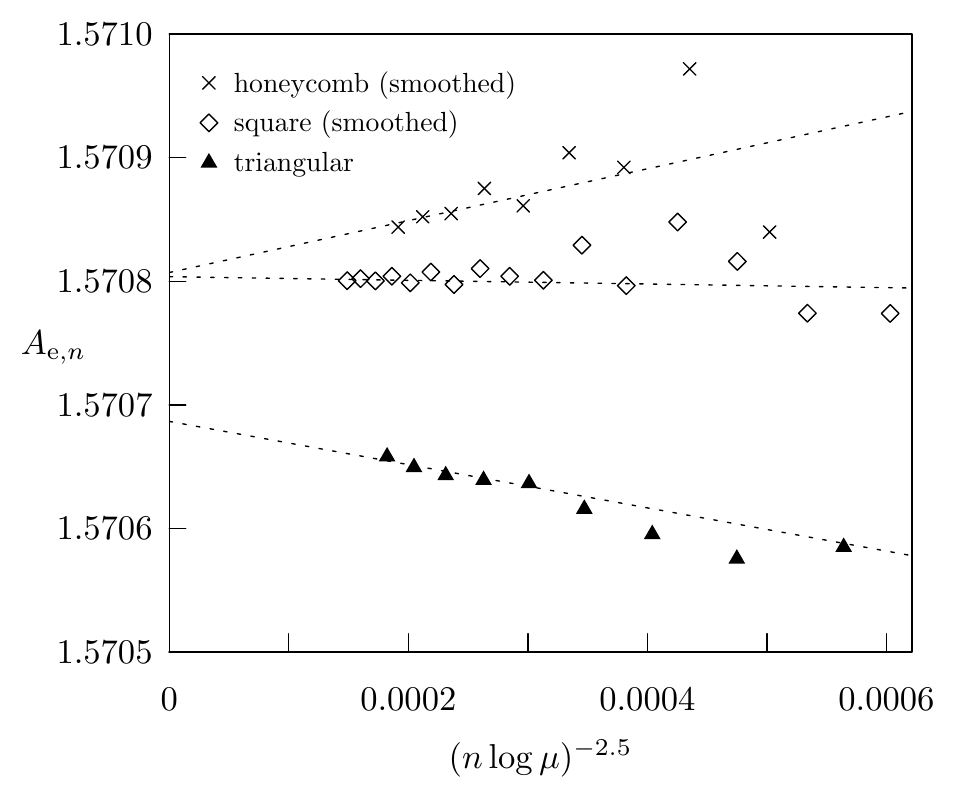}
\vspace{-3ex}
\end{center}
\caption{Direct fit estimates of $\Ae$ for two-dimensional lattices.
On the left is the result of fitting asymptotic form with no
correction to scaling, while on the right the leading correction to
scaling is fitted.\label{fig:Ae2_direct}}
\end{figure}

Assuming that $\gammae = 1$, the asymptotic form for $e_n$ is given by:
\begin{align}
e_n &\sim \Ae \mu^n n^{\gammae - 1}  
\left( 1 + \frac{a_1}{n} + \frac{a_2}{n^2} + \cdots 
+ \frac{b_0}{n^{\Delta_1}} + \frac{b_1}{n^{\Delta_2}} + \cdots \right),
\\ &\sim \Ae  \mu^n 
\left( 1 + \frac{a_1}{n} + \frac{a_2}{n^2} + \cdots 
+ \frac{b_0}{n^{\Delta_1}} + \frac{b_1}{n^{\Delta_2}} + \cdots \right).
\end{align}
For the square, honeycomb,  and triangular lattices we perform ``direct
fits'' to
\begin{align}
e_n / \mu^n &\sim \Ae + \left[ \frac{\Ae a_1}{n} + \frac{\Ae b_0}{n^{\Delta_1}} \right].
\end{align}
In this case the fits are trivial: one merely has to calculate the
ration $e_n / \mu^n$. For the union jack lattice, we fit
\begin{align}
\log e_n &\sim n \log \mu + \log \Ae + \left[\frac{a_1}{n} +
\frac{b_0}{n^{\Delta_1}} \right].
\end{align}
We plot our estimates against $1/n^{\Delta_1}$ in
Fig.~\ref{fig:Ae2_direct}, noting that once again this gives clear
linear trend which suggests that the analytic correction has zero
amplitude.

We now observe a remarkable fact: our amplitude estimates appear to be
converging towards the same constant. The coincidental convergence of
the estimates from series from four different lattices seems highly
unlikely, and thus we conclude that $\Ae$ is a universal quantity!

We test this conjecture further by \emph{assuming} that $a_0 = 0$, and
fitting 
\begin{align}
e_n / \mu^n &\sim \Ae + \frac{\Ae b_0}{n^{\Delta_1}} + \left[ \frac{\Ae
a_2}{n^2} + \frac{\Ae b_1}{n^{\Delta_2}} \right],
\end{align}
for the square, honeycomb, and triangular lattices only.  We plot the
resulting fits for $\Ae$ against the assumed form of the next-to-leading
non-analytic correction, i.e.  $1/n^{5/2}$~\cite{Caracciolo2005}. 
Although we expect that $a_2 = 0$ we do not rely upon this fact;
essentially the same estimates
are obtained for $\Ae$ from plots against $1/n^2$.

The resulting fits show a high degree of convergence. The scatter
between the different lattices is consistent with there being
non-trivial sub-leading corrections to scaling, which is not surprising
for such short series.

We estimate that $\Ae = 1.57075(10)$ for two-dimensional eSAW.  We note
in passing that $\pi/2 = 1.570796326794\cdots$ lies within the
confidence interval of our estimate, although we have no
theoretical basis for conjecturing that $\Ae = \pi/2$.

\begin{figure}[!htb]
\begin{center}
\includegraphics[scale=0.8]{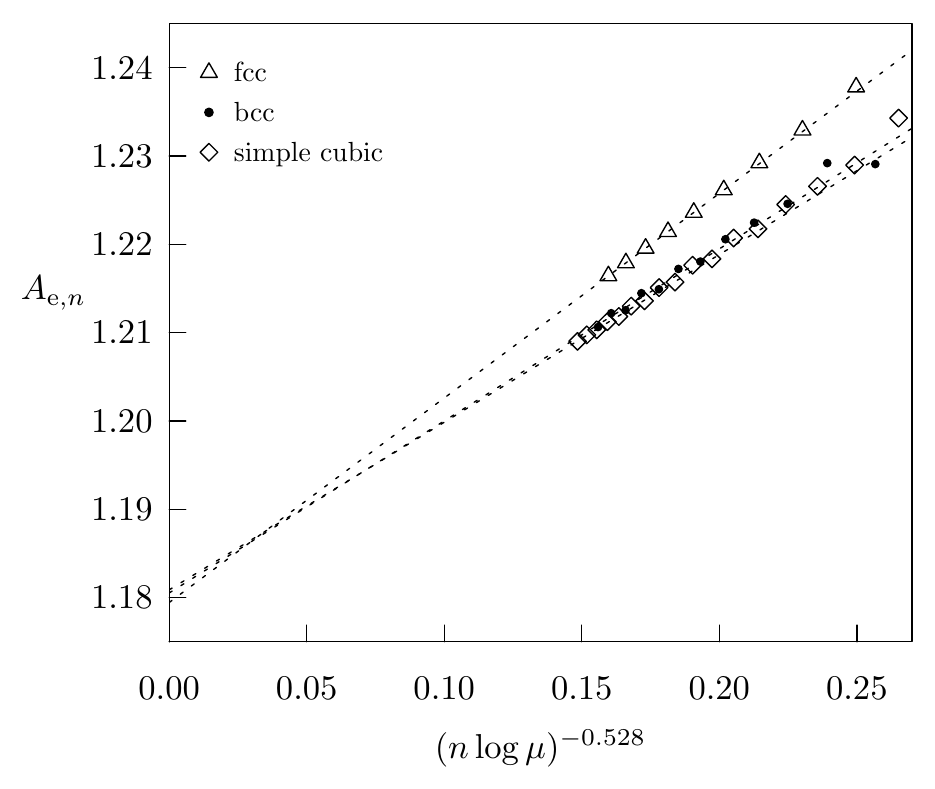}
\includegraphics[scale=0.8]{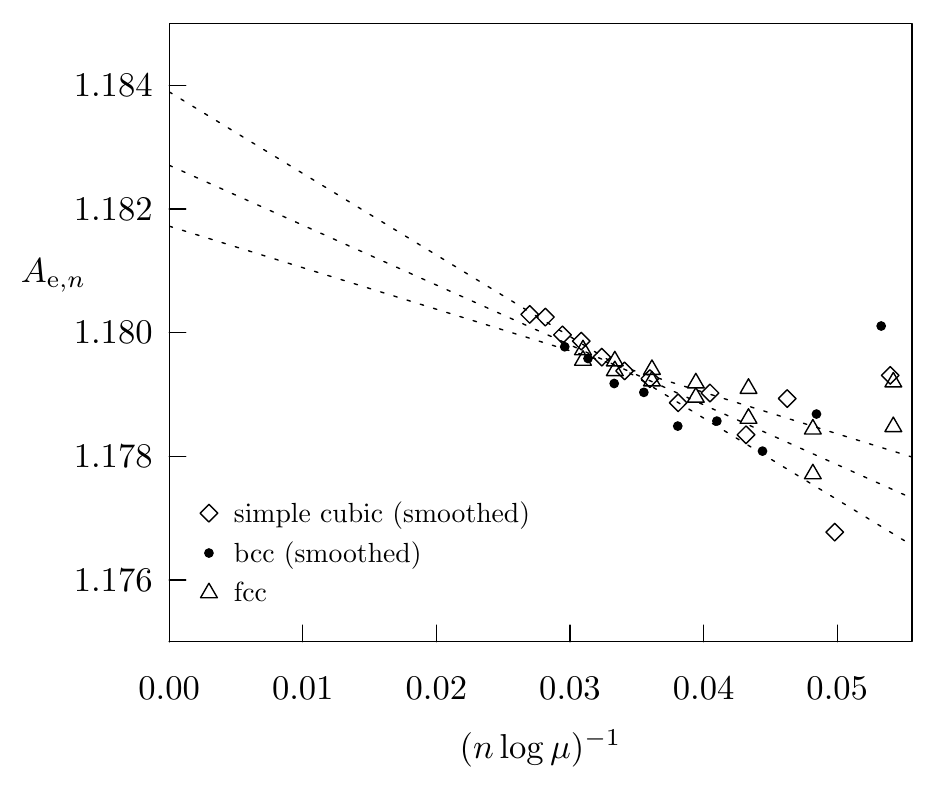}
\vspace{-3ex}
\end{center}
\caption{Direct fit estimates of $\Ae$ for three-dimensional lattices.
On the left is the result of fitting asymptotic form with no
correction to scaling, while on the right the leading correction to
scaling is fitted.\label{fig:Ae3_direct}}
\end{figure}

For the simple cubic, bcc, and fcc lattices 
we perform direct fits to
\begin{align}
e_n / \mu^n &\sim \Ae + \left[ \frac{\Ae a_1}{n} + \frac{\Ae
b_0}{n^{\Delta_1}} \right],\label{eq:a3a}
\end{align}
and
\begin{align}
e_n / \mu^n &\sim \Ae + \frac{\Ae b_0}{n^{\Delta_1}} + \left[ \frac{\Ae
a_1}{n} + \frac{\Ae b_1}{n^{\Delta_2}} +
\frac{\text{const.}}{n^{2\Delta_1}} \right].\label{eq:a3b}
\end{align}
For the simple cubic and bcc lattice we once again fit alternate values,
while for the fcc lattice we fit consecutive values.
We plot the resulting fits for $\Ae$ in Fig.~\ref{fig:Ae3_direct}.
We plot the fits from (\ref{eq:a3a}) against the leading neglected term,
$1/n^{\Delta_1}$.
For the fits from (\ref{eq:a3b}),
even if the analytic correction to scaling in 
has zero amplitude there are still two competing non-analytic
corrections with exponent in the vicinity of $-1$. Thus we plot fits
from
(\ref{eq:a3b}) against $1/n$.

Although we do not show these fits, we have tested the effect of using
values of $\Delta_1 = 0.528(12)$ within its error bar. In the fits from 
(\ref{eq:a3b}) this has the effect of extending the
extrapolations to the vicinity of 1.180 and 1.185.

The estimates for $\Ae$ from all three lattices once again
appear to be converging to a single value to good precision. This
strongly suggests that $\Ae$ is a universal quantity for
three-dimensional eSAW also.
We estimate that $\Ae = 1.183(3)$

\begin{figure}[!htb]
\begin{center}
\includegraphics[scale=0.8]{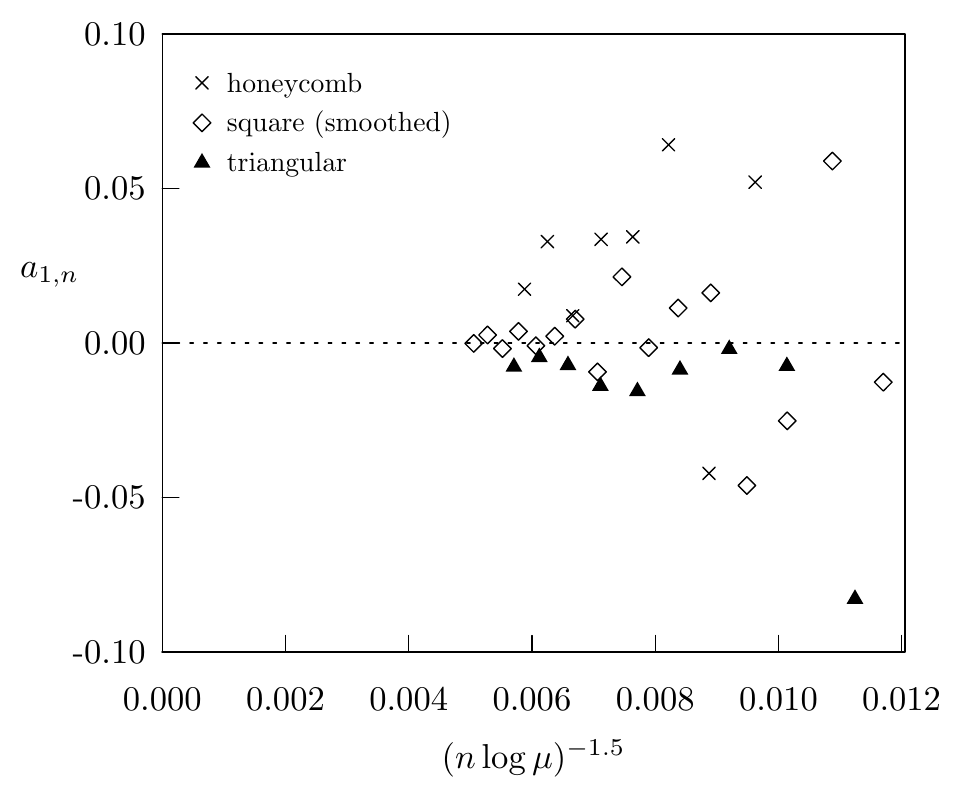}
\vspace{-3ex}
\end{center}
\caption{Direct fit estimates of the amplitude of the leading analytic
correction to scaling, $a_1$, for the two-dimensional lattices.\label{fig:a1_direct}}
\end{figure}

Finally, we will attempt to measure the amplitude of the leading
analytic correction to scaling for two-dimensional eSAW, which we expect
to be identically zero. To do so, we fit
\begin{align}
e_n / \mu^n &\sim \Ae + \frac{\Ae a_1}{n}+ \frac{\Ae b_0}{n^{\Delta_1}} + \left[ \frac{\Ae
a_2}{n^2} + \frac{\Ae b_1}{n^{\Delta_2}} \right],
\end{align}
for the square, honeycomb, and triangular lattices only.  
The inclusion of the $\Delta_1$ non-analytic correction is necessary to
be able to obtain sensible estimates for $a_1$.
We plot the
resulting estimates for $a_1$ from all three lattices in
Fig.~\ref{fig:a1_direct}.
We plot the fits against $1/n^{\Delta_2 - 1}$ since this is the relative
size of the first unfitted term (assuming that $a_2 = 0$).

We can see that for all three lattices the estimates for $a_1$ are in
the vicinity of zero, in agreement with our earlier observations based
on the trends of our fits. The honeycomb lattice fits are still quite
scattered, but the fits for the triangular and square lattices each
suggest that $a_1 = 0.00(2)$. Although this is perhaps not as strong
numerical evidence as that in favor of the universality of $\Ae$, we
nonetheless think that the evidence is convincing.

We do not attempt a similar analysis for the three-dimensional lattices,
because there is no realistic possibility of obtaining meaningful fits
for $a_1$. We discuss this further in Sec.~\ref{sec:corrections}.

\section{Discussion\label{sec:discussion}}

In the previous section, we confirmed that indeed $\gammae = 1$
and $\nue = \nu$ as we had predicted.

We now discuss our discovery that $\Ae$ appears to be universal in
Sec.~\ref{sec:amplitude}, the nature of corrections to scaling for
$e_n$ in 
Sec.~\ref{sec:corrections}, and the nature of the anti-ferromagnetic
singularity for eSAW in Sec.~\ref{sec:af}.
Then we mention some open questions for eSAW which will require further
study in Sec.~\ref{sec:open}, before discussing some possible
applications of eSAW in Sec.~\ref{sec:applications}.

\subsection{Universal amplitude\label{sec:amplitude}}

In Sec.~\ref{sec:amplitude_analysis} we presented strong numerical
evidence that the amplitude for the number growth of eSAW is a universal
quantity, with $\Ae = 1.57075(10)$ in two dimensions, and $\Ae =
1.183(3)$ in three dimensions. We had \emph{not} anticipated this
\emph{a priori}, unlike the case for our argument that $\gammae = 1$.
Indeed, the eSAW model was \emph{constructed} with the goal of creating
a model with $\gammae = 1$.  We do have a heuristic \emph{post hoc}
argument to explain why the amplitude is universal, which we present
below.

Once ends have been eliminated, $\mu$ is simply a measure of local
flexibility, and the polymer then only has one length scale, the Kuhn
length. Once this length has been rescaled we must be left with
a universal model, and since the length rescaling in $\Ae \mu^n$ occurs
completely independently of the amplitude, this means that $\Ae$ must be a
universal quantity.

We also observe that we now have an exact critical exponent
relation, $\gammae = 1$, to which there should be a corresponding
relation involving amplitudes.

It would be highly desirable to identify the critical amplitude with a
standard quantity in the scaling theory of polymers.

We note in passing that this universal amplitude may serve as a signal
of the $\theta$ transition of interacting self-avoiding walks. We know
that the amplitude of endless \emph{simple} random walks is unity.
For two-dimensional $\theta$-polymers we have $\gamma =
8/7$~\cite{Duplantier1987a}, and so we could expect that $\Ae$ for
endless $\theta$-polymers would have a distinct value, i.e. it should
lie between 1 and
$1.57\cdots$. The upper critical dimension for the $\theta$ transition
is $d = 3$, and so for three-dimensional endless $\theta$-polymers it is
likely that $\Ae = 1$.

We will just make one further point. Amplitudes for
lattice models, as opposed to amplitude ratios, are not typically
universal quantities.  For comparison, the amplitudes for SAW are $A =
1.1449355(5)$~\cite{Jensen2006} for the honeycomb lattice, and $A =
1.1770423(1)$~\cite{Jensen2004} for the square lattice.  In contrast,
the estimates for $\Ae$ from the square and honeycomb lattices agree to
the fourth decimal digit.

\subsection{Corrections to scaling\label{sec:corrections}}
The analysis for eSAW on two-dimensional lattices in
Sec.~\ref{sec:amplitude_analysis} 
supports our contention that there are no analytic corrections to
scaling
for the number growth, by confirming that the amplitude $a_1$
from (\ref{eq:enasympt}) is indistinguishable from zero for all
lattices considered.

With longer enumerations from the finite lattice method it may be
possible to estimate $a_2$ or even $a_3$, which could strengthen this
confirmation significantly. Monte Carlo simulations could also be used
to improve the accuracy of the estimation of $a_1$.

We find clear evidence that the leading non-analytic correction to
scaling for two-dimensional eSAW has exponent $-3/2$, which is the same
as for two-dimensional SAW~\cite{Caracciolo2005}. In particular, we do
not see any evidence of the correction to scaling with exponent
$-11/16$, predicted by CFT and observed for another recently invented
SAW variant~\cite{Gherardi2009,Gherardi2010}.

For three-dimensional eSAW it is certainly possible that either the
leading analytic correction to scaling has vanishing amplitude, or that
there are no analytic corrections to scaling at all.  However, it will
be far more difficult to gather numerical evidence to resolve this
question. This is because the leading non-analytic correction to scaling
dominates the analytic correction, and in addition there are \emph{two}
non-analytic corrections which would compete with the leading analytic
correction.  The leading analytic correction has exponent $-1$, while
the next-to-leading non-analytic corrections are $-2 \Delta_1 \approx
-1.06$ and $-\Delta_2$ which is believed to be close to $-1$.
Consequently we consider the prospects of obtaining sufficiently strong
numerical evidence to determine the amplitude of the leading analytic
correction to be remote for three-dimensional eSAW. 

Nonetheless, in the absence of a compelling argument for the existence
of confluent logarithmic corrections, we consider it likely that there
are no analytic corrections to scaling for three-dimensional eSAW.

\subsection{Anti-ferromagnetic singularity\label{sec:af}}

We have examined our series on the bipartite lattices - square, simple
cubic, and bcc - for any evidence of the anti-ferromagnetic (AF)
singularity (the honeycomb lattice is bipartite, but we are only
considering one of the sub lattices). The AF singularity arises from
polygon-like configurations: the odd-even effect is due to the fact that
self-intersections can only occur between either two odd or two even
sites of the walk.

The signature of the AF singularity for SAW is unambiguous for the same series lengths as
we consider here for eSAW. From differential approximant analyses one
can clearly detect the singularity on the negative real axis, at the
correct location. From direct fitting, when we include an AF term in our
fits we get clear signs of convergence to non-zero values for the fitted
amplitudes. Finally, the inverses of the generating functions of interest
provides a strong signal for the existence of the AF
singularity~\cite{Clisby2009}.

For the case of eSAW, none of these analyses are conclusive!

For the
bipartite lattice series, there is no clear signal from the DA analyses.
On occasion, the differential approximants have singularities on the
negative real axis, but the typical behavior is a strong signal of the
critical point on the positive axis, and then other singularities
scattered in the complex plane.

From the series themselves there are clearly oscillations, but the
period in no case appears to be strictly two, and we have not been able
to usefully apply the direct fitting approach.

Calculating the inverse of the generating functions, \`a la
\cite{Clisby2009}, leads to a strictly alternating series for SAW, but
for eSAW on the square lattice there is no regular pattern at all, while
for eSAW on the simple cubic and bcc lattices there is some hint that
the series are settling down to something akin to alternation, but this
is by no means definitive. Indeed, eSAW on the triangular lattice -
which certainly does not have an AF singularity - displays a
similar pattern to the simple cubic and bcc series.

In summary, the series analysis is inconclusive with respect to the
existence of the AF singularity, and it may well be that the observed
oscillations are determined by singularities in the complex plane.

We will now attempt to explain our intuition for whether there
\emph{should} be an AF singularity based on physical arguments.
This is purely speculative, and in the absence of further numerical
evidence should perhaps be given little weight.

For SAW, polygon-like configurations are the origin of the AF
singularity: the dominant contributing configurations consist of large
loops, and the odd-even effect arises from the interaction of the
end-points.

For eSAW, polygon-like configurations are exponentially rare, as
explained in Sec.~\ref{sec:open}. Large loops are possible, but the ends
are then forced apart by the endless condition.  Heuristically, for an
eSAW of $n$ steps, $O(n)$ steps can be part of a large loop, but then
$O(n^\nu)$ steps are required for the end-points to escape from each
other.  We expect that this will result in the odd-even effect from the
end-point interaction being exponentially suppressed or even eliminated
completely, and hence no AF singularity will lie on the radius of
convergence.  It may be that our intuition is incorrect and that merely
the amplitude of the AF singularity is suppressed, but our best guess is
that the AF singularity is completely absent for eSAW.

In future, we will revisit this question with longer series in hand in
an attempt to resolve this question definitively.

\subsection{Open questions\label{sec:open}}

One important topic for eSAW which we have not yet addressed is the
behavior of eSAW with fixed end-points, and in particular the
asymptotic behavior of $e_n(x)$ as $n$ approaches infinity with $x$
fixed. For SAW, it is well known that
\begin{align}
c_n(x) \sim B(x) \mu^n n^{2-\alpha},
\end{align}
but the situation for eSAW is radically different. For eSAW in $d$
dimensions, in the limit of large $n$ it seems that the eSAW must become
quasi $(d-1)$ dimensional objects, with the eSAW effectively confined to a
`slab' with dimensions fixed by the end-to-end vector $x$. This slab could
have quite a complicated structure, with the parallel surfaces allowed
to have arbitrary shape provided that they are always separated by $x$.
Thus it is not clear if the resulting eSAW would have growth constant of
$d-1$ dimensional SAW or a value that is intermediate between $d-1$ and
$d$ dimensional SAW. The challenge is to
understand and possibly predict the values of $\kappa$, $a$, and $\Be$ in
\begin{align}
e_n(x) \sim \Be(x) \kappa^n n^{a}.
\end{align}
If eSAW in this limit are \emph{exactly} like fixed end-point SAW in
$d-1$ dimensions, then we would expect that $\kappa = \mu_{d-1}$ and $a
= -(d-1)\nu_{d-1}$.  This is a question that might be resolvable with a
good theoretical argument, and it is also clearly desirable to study
this question numerically. 

Another important question is whether the pivot
algorithm~\cite{Lal1969,Madras1988} is ergodic for eSAW.
One reason for the importance of this question is that recent
improvements to the implementation of the pivot
algorithm~\cite{Kennedy2002a,Clisby2010,Clisby2010a} have made it even
more powerful. In addition, it is now clear that the pivot algorithm can
be used for extremely accurate calculations of $\mu$, $\gamma$, and
$\nu$ for SAW.

We expect that the pivot algorithm is indeed ergodic, since eSAW are
very similar in nature to SAW, and in some sense they allow for less
pathological behavior with respect to the pivot algorithm than SAW. This
is true in the following sense: for SAW it is possible to form spiral
structures with their ends which can only be unfurled extremely slowly
with pivot moves. In contrast, for eSAW it is never possible for an end
to become trapped, and thus it would seem that `most' sites of a typical
eSAW should be candidates for valid pivot moves. The key question is
whether there are novel equivalence classes of eSAW which are invariant
under pivot moves, which would result in non-ergodicity.

It may be difficult to prove that the pivot algorithm is indeed ergodic
for eSAW. The difficulty lies in the fact that the non-local
interactions change when a pivot is made, as the end-to-end vector
typically changes after a successful move. Combining fixed end-point pivot
moves~\cite{Madras1990,JansevanRensburg1990} with
standard pivot moves might allow for a more straightforward proof, but
this is a sub-optimal solution because fixed end-point moves are less
efficient for the purpose of simulation.
Application of the pivot algorithm to eSAW has not been the focus here,
but it is a topic we intend to explore in the future.

Finally, much remains to be done regarding the application of theoretical
techniques to eSAW. In particular, the renormalization group in
conformation space~\cite{Oono1981} could be adapted to the study of
eSAW, and should provide a strong theoretical argument that $\gammae = 1$
in all dimensions. 

\subsection{Applications\label{sec:applications}}

Perhaps the application for which eSAW has the clearest advantage over
the competing models of SAP and SAW is in the study of polymer knotting.
Simulations for eSAW promise to be far more rapid than
for SAP, provided the pivot algorithm with free end-point pivot moves can
be proved to be ergodic (see Sec.~\ref{sec:open}). 

At first glance, the eSAW series seem to hold less information than the
corresponding SAW series of the same lengths, but it is quite possible
that this is a short length effect, and that eSAW becomes superior to SAW
at longer lengths. 
The absence of analytic corrections-to-scaling, and the possibility of
deriving a theoretical prediction for the universal amplitude $\Ae$, are
significant advantages for eSAW over SAW.
The large $n$ limit can be probed via finite lattice
method enumerations (perhaps approaching $n = 80$ or so for the square
lattice), and also via Monte Carlo simulation using the pivot
algorithm (where $n \gtrsim 10^8$).
Thus we believe that eSAW hold considerable promise for allowing extremely
accurate calculations for critical exponents and the connective
constant for polymers in two and three dimensions.

Finally, we speculate that eSAW may be useful in the context of exact
results on the honeycomb lattice~\cite{Duminil-Copin2012b,Beaton2012c}.
eSAW are a naturally defined special subset of SAW which have zero winding
angle.

\subsection{Generalizations}

The concatenation idea of eSAW can be straightforwardly applied to other
models of linear polymers, including interacting self-avoiding
walks (ISAW)~\cite{Orr1947}, the Domb-Joyce model of weakly interacting
walks~\cite{Domb1972}, and indeed continuum models such as bead-spring
models. For these models, the interaction energy would need to be
converted to an interaction energy per unit length, but other than that
there are no significant complications.
The self-avoiding trail
model~\cite{Malakis1975,Malakis1976,Guttmann1985} and generalizations
(e.g. \cite{Bedini2012}) could also be easily made `endless'.

One interesting consequence of the existence of a universal amplitude
for the endless Domb-Joyce model, is that as the interaction parameter is
varied, the (weighted) growth constant will vary while the
amplitude will remain fixed. i.e. we would expect that for interaction
strength $J$:
\begin{align}
e_n(J) &\sim \Ae \mu(J)^n.
\end{align}
Of course this exactly what happens for $\mu(J)$ and $\gamma$ for the usual
Domb-Joyce model.

It would be interesting to see if the eSAW idea could be extended to the
continuum Edwards model~\cite{Edwards1965}.

As flagged in Sec.~\ref{sec:honeycomb}, it may be possible to allow for
eSAW with an odd number of steps on the honeycomb lattice. More
generally, the rule for concatenation of a walk with itself could be
expanded to allow for symmetry operations such as reflections to be
applied.

One weakness of the eSAW model is that there is no natural way to
incrementally grow an eSAW. For a SAW, any sub-walk is also a SAW, but
this is not true of eSAW. Interestingly, the set of eSAW for which any
sub-walk is also an eSAW is precisely the set of directed walks. This
can be seen straightforwardly by noting that: (a) If an eSAW has two
steps in opposite directions, then the sub-walk which begins and ends
with these steps is not an eSAW, and (b) Every directed walk is an eSAW.
Perhaps there are other subsets of eSAW which may be of more interest.

\begin{figure}[htb]
\begin{center}
\begin{tikzpicture}[ultra thick,scale=0.4]
\begin{scope}[shift={(0,0)}]
\sww \nwa \ew 
\sww \nwb \ew 
\sww \nwc \ew 
\end{scope}
\begin{scope}[shift={(6,-2)}]
\sdw \nwa \ew 
\sdw \nwb \ew 
\sdw \nwc \ew 
\end{scope}
\begin{scope}[shift={(5,3)}]
\sdw \nwa \ew 
\sdw \nwb \ew 
\sdw \nwc \ew 
\end{scope}
\begin{scope}[shift={(1,-5)}]
\sdw \nwa \ew 
\sdw \nwb \ew 
\sdw \nwc \ew 
\end{scope}
\begin{scope}[shift={(-5,-3)}]
\sdw \nwa \ew 
\sdw \nwb \ew 
\sdw \nwc \ew 
\end{scope}
\begin{scope}[shift={(-1,5)}]
\sdw \nwa \ew 
\sdw \nwb \ew 
\sdw \nwc \ew 
\end{scope}
\begin{scope}[shift={(-6,2)}]
\sdw \nwa \ew 
\sdw \nwb \ew 
\sdw \nwc \ew 
\end{scope}
\end{tikzpicture}
\end{center}
\caption{An example of a self-avoiding network created via concatenation
of a star polymer on the square lattice.\label{fig:network}}
\end{figure}
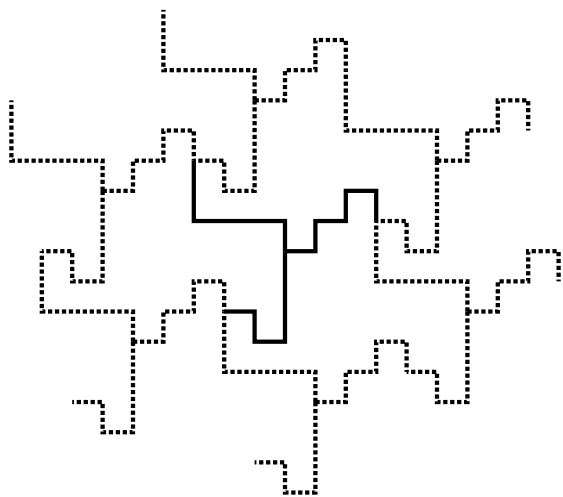

It may be possible to study polymer networks using variations of eSAW,
by concatenating together repeated copies of star or branched polymers.
An example of a two-dimensional network created by such a recipe is
shown in Fig.~\ref{fig:network}.

\section{Conclusion\label{sec:conclusion}}

We have introduced a new model of self-avoiding walks for which we have
theoretical and numerical evidence that end-effects should be completely
absent.

The enumeration results and analysis confirmed our expectations that
$\gamma = 1$ for eSAW, but more
serious numerical work should be done to provide stronger confirmation.
In future we will apply Monte Carlo
techniques~\cite{Madras1988,Kennedy2002a,Clisby2010,Clisby2010a} to
accurately estimate critical exponents.
It would be desirable to improve the rigor
of our heuristic arguments that 
$\gammae = 1$ and $\nue = \nu$.

We have made the surprising discovery that the eSAW model has a
universal amplitude, and estimate that $\Ae = 1.57075(10)$ in two
dimensions, and $\Ae = 1.183(3)$ in three dimensions.  We have given one
argument for this phenomenon, but it would be highly desirable to find
other theoretical justifications.  It remains to be seen if other models
can be found which possess universal amplitudes.  We also 
believe that there are no analytic corrections to scaling for the number
growth of eSAW, and
have supporting numerical
evidence for two-dimensional eSAW.

The concatenation technique described in this paper can be
straightforwardly applied to
other models of polymers to create new `endless' models. We are hopeful
that the simple idea introduced here will open up a fruitful avenue of
research.

\FloatBarrier

\appendix

\section{Enumeration data\label{sec:data}}
\begin{center}
\footnotesize
\begin{tabular}{rrrrr}
\hline
$n$ & $e_n$ & $er_n$ & $i_n$ & $d_n$ \\
\hline
1 & 4 & 4 & 4 & 2 \\
2 & 12 & 32 & 8 & 2 \\
3 & 28 & 156 & 24 & 4 \\
4 & 76 & 640 & 64 & 8 \\
5 & 204 & 2380 & 200 & 20 \\
6 & 540 & 8304 & 504 & 42 \\
7 & 1404 & 27580 & 1400 & 100 \\
8 & 3724 & 89216 & 3648 & 228 \\
9 & 9748 & 280980 & 9720 & 540 \\
10 & 25772 & 869360 & 25560 & 1278 \\
11 & 67940 & 2649284 & 67936 & 3088 \\
12 & 179068 & 7967808 & 178464 & 7436 \\
13 & 472580 & 23722244 & 472576 & 18176 \\
14 & 1245620 & 69977040 & 1244208 & 44436 \\
15 & 3286308 & 204845220 & 3286080 & 109536 \\
16 & 8666956 & 595628032 & 8663232 & 270726 \\
17 & 22861604 & 1721605028 & 22861600 & 672400 \\
18 & 60301764 & 4950241488 & 60291504 & 1674764 \\
19 & 159063748 & 14167239268 & 159063744 & 4185888 \\
20 & 419582396 & 40376066560 & 419556560 & 10488914 \\
21 & 1106796012 & 114635882508 & 1106794584 & 26352252 \\
22 & 2919664716 & 324361357936 & 2919596768 & 66354472 \\
23 & 7701735292 & 914912704060 & 7701735288 & 167429028 \\
24 & 20317148476 & 2573307218304 & 20316965760 & 423270120 \\
25 & 53595401604 & 7218778025700 & 53595401400 & 1071908028 \\
26 & 141385319932 & 20201661701936 & 141384847344 & 2718939372 \\
27 & 372973583908 & 56408010198948 & 372973574160 & 6906918040 \\
28 & 983913216292 & 157179257119360 & 983911970608 & 17569856618 \\
29 & 2595592842460 & 437133635653148 & 2595592842456 & 44751600732 \\
30 & 6847276871420 & 1213539966457200 & 6847273559040 & 114121225984 \\
31 & 18063481834652 & 3363306014128316 & 18063481834648 & 291346481204 \\
32 & 47652479932748 & 9306731807696384 & 47652471265792 & 744569863528 \\
33 & 125710536151052 & 25715129922994380 & 125710536083088 & 1904705092168 \\
34 & 331633319500804 & 70954363887188496 & 331633296639192 & 4876960244694 \\
35 & 874875502020444 & 195525476447120540 & 874875502018840 & 12498221457412 \\
\hline
\end{tabular}
\nopagebreak
\normalsize
\captionof{table}{Enumeration data for the square lattice.\label{tab:z2}}
\end{center}

\begin{center}
\small
\begin{tabular}{rrrrr}
\hline
$n$ & $e_n$ & $er_n$ & $i_n$ & $d_n$ \\
\hline
1 & 6 & 6 & 6 & 3 \\
2 & 30 & 72 & 24 & 6 \\
3 & 126 & 558 & 120 & 20 \\
4 & 606 & 3744 & 576 & 72 \\
5 & 2766 & 23070 & 2760 & 276 \\
6 & 13134 & 135288 & 12984 & 1082 \\
7 & 60990 & 764862 & 60984 & 4356 \\
8 & 286014 & 4214784 & 285408 & 17838 \\
9 & 1333926 & 22773798 & 1333800 & 74100 \\
10 & 6235950 & 121158840 & 6233160 & 311658 \\
11 & 29160390 & 636821526 & 29160384 & 1325472 \\
12 & 136280046 & 3312601632 & 136266336 & 5677764 \\
13 & 637801014 & 17087666310 & 637801008 & 24530808 \\
14 & 2981709558 & 87499403544 & 2981648544 & 106487448 \\
15 & 13958158806 & 445295758230 & 13958155920 & 465271864 \\
16 & 65278026174 & 2253803314176 & 65277740160 & 2039929380 \\
17 & 305608333062 & 11353558220742 & 305608333056 & 8988480384 \\
18 & 1429710254742 & 56952225122616 & 1429708907808 & 39714136328 \\
19 & 6693530115990 & 284620991825574 & 6693530115984 & 176145529368 \\
20 & 31322452467006 & 1417625514526560 & 31322446230480 & 783061155762 \\
21 & 146644481901582 & 7039475232292062 & 146644481840472 & 3491535281916 \\
22 & 686366924272686 & 34860099689171064 & 686366895112272 & 15599247616188 \\
23 & 3213444320539710 & 172198320473193822 & 3213444320539704 & 69857485229124 \\
24 & 15042778398965358 & 848672945442198144 & 15042778262399904 & 313391213799998 \\
\hline
\end{tabular}
\normalsize
\captionof{table}{Enumeration data for the simple cubic lattice.\label{tab:z3}}
\end{center}

\begin{center}
\footnotesize
\begin{tabular}{rrrrrr}
\hline
$n$ & honeycomb ($e_{2n}$) & triangular & union jack & bcc & fcc
 \\
\hline
1 & 6 & 6 & 8 & 8 & 12 \\
2 & 18 & 30 & 56 & 56 & 132 \\
3 & 60 & 114 & 296 & 344 & 1284 \\
4 & 210 & 486 & 1816 & 2312 & 12756 \\
5 & 726 & 1986 & 10728 & 14648 & 126612 \\
6 & 2448 & 8202 & 62648 & 96152 & 1265604 \\
7 & 8448 & 33858 & 365464 & 622616 & 12652932 \\
8 & 28818 & 140262 & 2130488 & 4066376 & 126616404 \\
9 & 98556 & 580938 & 12424112 & 26456024 & 1267690812 \\
10 & 336618 & 2407710 & 72464136 & 172470776 & 12697464132 \\
11 & 1150320 & 9980394 & 422680024 & 1124590904 & 127218134748 \\
12 & 3928944 & 41384106 & 2465699704 & 7328945336 & 1274923451364 \\
13 & 13419204 & 171634482 & 14384367304 & 47829176312 & 12779117065740 \\
14 & 45828192 & 711932538 & 83919450272 & 311758824440 & 128110554439020 \\
15 & 156512220 & 2953394214 & 489611186736 & 2035191240584 & \\
16 & 534463698 & 12253006566 & 2856635570680 & 13269735071816 & \\
17 & 1825120584 & 50838740046 & 16667509269248 & 86634355605416 & \\
18 & 6232259412 & 210946278126 & 97251708586688 & 565037351668904& \\
19 & 21281168202 & 875324356398 & & & \\
20 & 72666555570 & 3632308990686 & & & \\
21 & 248124503652 & 15073375131306 & & & \\
22 & 847224827676 & 62553271396074 & & & \\
23 & 2892836367066 & & & & \\
24 & 9877456541376 & & & & \\
25 & 33725891989626 & & & & \\
\hline
\end{tabular}
\normalsize
\captionof{table}{Enumeration data for the honeycomb, triangular, union
jack, bcc, and fcc lattices.\label{tab:honeycomb}}
\end{center}

\section*{Acknowledgments}
I would like to thank Tony Guttmann for helpful discussions which led me
to consider the anti-ferromagnetic singularity more deeply, and for
comments on the paper. I would like to thank Mireille
Bousquet-M\'{e}lou for comments which led me to consider the
relationship between cyclic permutations of bridges and endless
self-avoiding walks. Financial support from the ARC Centre of Excellence
for Mathematics and Statistics of Complex Systems is gratefully
acknowledged.

\end{document}